\documentclass[12pt]{iopart}

\usepackage{cite}
\usepackage{psfrag}
\usepackage[mathcal]{euscript}
\usepackage{graphicx}

\eqnobysec

\newcommand{\eqref}[1]{(\ref{eq:#1})}
\newcommand{\apxref}[1]{\ref{apx:#1}}
\newcommand{\secref}[1]{section~\ref{sec:#1}}
\newcommand{\secreftwo}[2]{sections~\ref{sec:#1} and~\ref{sec:#2}}
\newcommand{\Secref}[1]{Section~\ref{sec:#1}}
\newcommand{\figref}[1]{figure~\ref{fig:#1}}
\newcommand{\Figref}[1]{Figure~\ref{fig:#1}}

\newcommand{\Tabref}[1]{Table~\ref{table:#1}}

\newcommand{\mathnotation}[2]{\newcommand{#1}{\ensuremath{#2}}}

\renewcommand{\l}{\left}			
\renewcommand{\r}{\right}			
\mathnotation{\pd}{\partial}			
\mathnotation{\ldef}{\mathrel{\raisebox{.069ex}{:}\!\!=}}
\mathnotation{\rdef}{\mathrel{=\!\!\raisebox{.069ex}{:}}}
\mathnotation{\half}{{\textstyle {1\over2}}}	
\mathnotation{\third}{{\textstyle {1\over3}}}	
\mathnotation{\grad}{\nabla}			
\mathnotation{\curl}{\grad\times}		
\renewcommand{\div}{\grad\cdot}			
\mathnotation{\lapl}{\nabla^2}			
\mathnotation{\dint}{\,{\mathrm{d}}}		
\newcommand{\Order}[1]{\ensuremath{\Or\!\l(#1\r)}}
\renewcommand{\time}{t}				
\mathnotation{\lyapexp}{\lambda}		
\mathnotation{\lyapexpinf}{\lyapexp^\infty}	
\mathnotation{\x}{x}				
\mathnotation{\xv}{{\bi{\x}}}			
\mathnotation{\velc}{v}				
\mathnotation{\vel}{{\bi{\velc}}}		
\mathnotation{\dvdx}{G}				
\mathnotation{\dvdxhat}{\widehat{\dvdx}}	
\mathnotation{\dvdxsym}{\widehat{A}}		
\mathnotation{\Jacm}{M}				
\mathnotation{\Hesm}{K}				
\mathnotation{\Hesmh}{\widehat{K}}		
\mathnotation{\sdim}{n}				
\mathnotation{\ncontr}{m}			
\mathnotation{\NI}{N_{\mathrm{I}}}		
\mathnotation{\NII}{N_{\mathrm{II}}}		
\mathnotation{\edir}{e}				
\mathnotation{\ediru}{\hat{\edir}}		
\mathnotation{\ediruv}{{\mathbf{\ediru}}}
\mathnotation{\ediruinf}{\ediru^\infty}
\mathnotation{\ediruvinf}{\mathbf{\ediru}^\infty}
\mathnotation{\lagrc}{a}			
\mathnotation{\lagrcv}{{\bi{\lagrc}}}		
\mathnotation{\metric}{g}			
\mathnotation{\metricx}{h}			
\mathnotation{\detmetric}{|\metric|}		
\mathnotation{\detmetricx}{|\metricx|}		
\mathnotation{\nugr}{\Lambda}			
\mathnotation{\nudec}{\gamma}			
\mathnotation{\gradlc}{\grad_0}			
\mathnotation{\curllc}{\gradlc\times}		
\mathnotation{\divlc}{\gradlc\cdot}		
\newcommand{\transp}[1]{{#1}^T}			

\mathnotation{\Usvd}{U}				
\mathnotation{\Vsvd}{V}				
\mathnotation{\Fsvd}{F}				%

\mathnotation{\Vgrnu}{\Psi}			
\mathnotation{\VgrU}{\Phi}			
\mathnotation{\VgrV}{\Theta}			
\mathnotation{\Vgrdetg}{\Omega}			
\mathnotation{\X}{X}				
\mathnotation{\Xh}{\widehat{\X}}		

\mathnotation{\Vgrnut}{\widetilde{\Vgrnu}}
\mathnotation{\VgrUt}{\widetilde{\VgrU}}
\mathnotation{\VgrVt}{\widetilde{\VgrV}}
\mathnotation{\Vgrdetgt}{\widetilde{\Vgrdetg}}
\mathnotation{\Vgrnuinf}{\Vgrnu^\infty}
\mathnotation{\VgrVinf}{\VgrV^\infty}
\mathnotation{\Vgrdetginf}{\Vgrdetg^\infty}
\mathnotation{\homo}{{\mathrm{h}}}		
\mathnotation{\Vgrnudrive}{\Vgrnu^{\mathrm{drive}}}
\mathnotation{\VgrVdrive}{\VgrV^{\mathrm{drive}}}
\mathnotation{\Hesmt}{\widetilde{\Hesm}}
\mathnotation{\Vsvdt}{\widetilde{\Vsvd}}
\mathnotation{\Vsvdinf}{\Vsvd^\infty}

\mathnotation{\Qqr}{Q}				
\mathnotation{\Rqr}{R}				
\mathnotation{\rqr}{r}				
\mathnotation{\nuqr}{\Delta}			
\mathnotation{\Dqr}{D}				
\mathnotation{\DDqr}{{\mathcal{F}}}		
\mathnotation{\Gqr}{\bar{G}}			
\mathnotation{\Aqr}{\bar{A}}			
\mathnotation{\GSqr}{d}				
\mathnotation{\Wqr}{W}				
\mathnotation{\WgrQ}{\phi}			
\mathnotation{\Wgrnu}{\psi}			
\mathnotation{\Wgrr}{\xi}			
\mathnotation{\WgrW}{\theta}			
\mathnotation{\Wgra}{\eta}			
\mathnotation{\Xt}{\bar{\X}}			
\mathnotation{\Yt}{Y}				

\mathnotation{\Ricci}{R}			
\mathnotation{\Rcurv}{R}			
\mathnotation{\Riccirotc}{\omega}		

\mathnotation{\Eul}{E}				
\mathnotation{\Ugrnu}{\Psi^\Eul}		
\mathnotation{\UgrU}{\Phi^\Eul}			
\mathnotation{\UgrV}{\Theta^\Eul}		
\mathnotation{\Ugrnut}{\widetilde{\Psi}^\Eul}
\mathnotation{\UgrUt}{\widetilde{\Phi}^\Eul}
\mathnotation{\UgrVt}{\widetilde{\Theta}^\Eul}
\mathnotation{\Ugrnuinf}{\Psi_{\kappa\nu}^{\Eul\infty}}
\mathnotation{\UgrUinf}{\Phi_{\kappa\mu\nu}^{\Eul\infty}}

\def\QR{$QR$}
\def\SVD{SVD}
\def\ABC{$ABC$}

\begin{document}

\paper[Derivatives and Constraints in Chaotic Flows]{Derivatives and
	Constraints in Chaotic Flows:\\ Asymptotic Behaviour and a Numerical
	Method} \author{Jean-Luc Thiffeault}

\address{Department of Applied Physics and Applied Mathematics\\
	Columbia University, New York, NY 10027, USA}

\ead{jeanluc@mailaps.org}

\begin{abstract}

In a smooth flow, the leading-order response of trajectories to infinitesimal
perturbations in their initial conditions is described by the finite-time
Lyapunov exponents and associated characteristic directions of stretching.  We
give a description of the second-order response to perturbations in terms of
\emph{Lagrangian derivatives} of the exponents and characteristic directions.
These derivatives are related to generalised Lyapunov exponents, which
describe deformations of phase-space elements beyond ellipsoidal.  When the
flow is chaotic, care must be taken in evaluating the derivatives because of
the exponential discrepancy in scale along the different characteristic
directions.  Two matrix decomposition methods are used to isolate the
directions of stretching, the first appropriate in finding the asymptotic
behaviour of the derivatives analytically, the second better suited to
numerical evaluation.  The derivatives are shown to satisfy differential
constraints that are realised with exponential accuracy in time.  With a
suitable reinterpretation, the results of the paper are shown to apply to the
Eulerian framework as well.

\end{abstract}

\submitto{Physica D}


\section{Introduction}
\label{sec:intro}

We consider a collection of coupled ordinary differential equations (ODEs)
associated with a vector field~\cite{GuckHolmes} (a continuous-time dynamical
system).  The general solution to these equations defines a \emph{flow}---a
mapping from the phase-space domain onto itself.\footnote{In the mathematics
literature the term \emph{flow} is reserved for autonomous systems, the term
\emph{transformation} being preferred for the solutions generated by
time-dependent vector fields~\cite[p.~96]{ArnoldODE}.  Because of the
compelling nature of the fluid analogy, \emph{flow} will be used in the more
general sense in this paper.}  For smooth vector fields, the flow can be
regarded as a smooth coordinate transformation (a diffeomorphism) from the set
of initial conditions to the state of the system at some later
time~\cite[p.~276]{ArnoldODE}.  The coordinates describing the initial
conditions are called the \emph{Lagrangian coordinates}, and those describing
the state at a later time are called the \emph{Eulerian coordinates}.

Generally, even a smooth vector field will lead to chaotic
dynamics,\footnote{Important exceptions are 1D vector fields, and 2D
autonomous vector fields~\cite{Eckmann1985}.  There are also other, more
restricted classes of nonchaotic vector fields.} with trajectories of nearby
phase-space elements diverging rapidly from each other, at an average rate
close to exponential for long times.  The allowable rates are called the
\emph{Lyapunov exponent} of the flow~\cite{Lyapunov}, and are associated with
\emph{characteristic directions} of stretching.  For chaotic flows, the
transformation to Lagrangian coordinates becomes exceedingly contorted, and in
practise it can no longer be inverted, due the exponentially growing errors on
the position of phase-space particles.  Nevertheless, the presence of chaos
can actually be advantageous because it leads to a large separation of
timescales along the different characteristic directions of the flow.  This
separation of scale was used by Boozer~\cite{Boozer1992}, Tang and
Boozer~\cite{Tang1996,Tang2000}, and Giona and
Adrover~\cite{Giona1998,Adrover1999} to study fluid mixing and the dynamo
problem.

Many equations of fluid dynamics are ``advective'' in nature, in that they
describe the motion of a scalar or vector field as it is dragged by a flow,
and possibly influenced by other effects such as diffusion and sources.
Examples are the scalar advection--diffusion equation~\cite{LandauFluids} and
the induction equation for a magnetic field~\cite{STF}.  When expressed in
Lagrangian coordinates, the advective term drops out of these types of
equations.  For scalar and vector advection--diffusion equations, one is left
with a diffusion equation with anisotropic diffusivity.  The anisotropic
diffusivity arises because the Jacobian of the transformation between
Lagrangian and Eulerian coordinates is not orthogonal.  In a chaotic flow, the
complexity of the transformation leads to a singular Jacobian, which is
reflected in an exponentially-growing diffusivity that enhances
mixing~\cite{Tang1996,Thiffeault2001f}.

In Refs.~\cite{Tang1996,Thiffeault2001f}, the singular Jacobian is decomposed
to isolate the dominant direction of enhanced diffusion, leading to an
expression in terms of the finite-time Lyapunov exponents and characteristic
directions of stretching.  To get a full solution of the advection--diffusion
problem, the Lagrangian derivatives of the finite-time Lyapunov exponents and
of the characteristic directions are needed.  This is because the Lagrangian
coordinate frame is position-dependent, so when derivatives of vector fields
are taken one must also differentiate the basis vectors themselves (this is
the same procedure as in covariant differentiation, or when fictitious forces
appear after transforming to a rotating frame).  Thus the necessity of
obtaining Lagrangian derivatives of the vectors defining the coordinate frame.
Because Lagrangian coordinates are also \emph{stretched} with respect to
Eulerian coordinates, the derivatives of the characteristic rates of
separations (as characterised by the finite-time Lyapunov exponents) are also
needed.

The problem of finding the asymptotic form of Lagrangian derivatives of the
coordinate transformation induced by a flow has been addressed previously by
Dressler and Farmer~\cite{Dressler1992} and Taylor~\cite{Taylor1993} in a
different context.  They examined the asymptotic behaviour of the Hessian, the
quadratic form consisting of the second derivatives of the flow.  The Hessian
is the term that follows the Jacobian matrix in a Taylor expansion of the
coordinate transformation form Lagrangian to Eulerian coordinates.  Dressler
and Farmer call the growth rates of the Hessian \emph{generalised} or
\emph{higher-order} Lyapunov exponents.  Their motivation lay in
characterising the growth of nonlinear distortions of geometric quantities
evolving under the influence of one-dimensional maps.  The Lyapunov exponents
quantify the leading order stretching of an infinitesimal ellipse moving with
the phase fluid, and the generalised exponents describe deviations from an
elliptical shape.  Dressler and Farmer provided numerical results only for the
largest Lyapunov exponent, because no numerical method exist to evolve the
Hessian in a numerically stable manner that is not susceptible to limited
precision.  We provide such a method here.

In the Eulerian picture, the higher-order exponents also characterise the
growth of extrinsic curvature of curves and surfaces embedded in a flow
(material lines and surfaces), and so can be connected to work describing the
evolution of the curvature of material lines in turbulent flows that
originated with
Batchelor~\cite{Batchelor1952,Pope1988,Pope1989,Girimaji1991,Drummond1991},
and more recently was applied to chaotic
flows~\cite{Liu1996,Hobbs1998,Cerbelli2000}.  Our results---derived in the
Lagrangian frame---can easily be adapted to the Eulerian picture, and provide
a comprehensive view of second-order deformation processes in chaotic flows.
The connexion between the Eulerian and Lagrangian pictures is made in an
appendix.

Our approach is similar to Refs.~\cite{Dressler1992,Taylor1993} but is more
general and applies to flows rather than maps: we aim to give estimates of the
asymptotic growth rates of the Lagrangian derivatives of finite-time Lyapunov
exponents and characteristic directions by appealing to arguments of
``genericity'' of the quantities involved, thus showing that the estimates
will hold in essiantially all cases.  These arguments are formal in nature, in
the sense that they do not provide a mathematical proof of the results, since
there will always exist flows that can be specifically chosen to violate any
of the estimates.  In particular, there could be flows with degenerate
Lyapunov exponents (finite-time Lyapunov exponents that are degenerate for
short periods of time do not concern us).  We expect that even for degenerate
exponents the results of the paper are applicable in a limited form.  In
practise the asymptotic behaviour holds to great accuracy for all flows
examined.  This will be verified in the numerical section of the paper.

To obtain the estimates of asymptotic behaviour of derivatives, we perform a
singular value decomposition (\SVD) of the tangent mapping of the flow, and
differentiate the ODEs derived by Greene and Kim~\cite{Greene1987} directly.
A careful analysis of the equations, with the assumption of a nondegenerate
spectrum and a bounded attractor as in Goldhirsch \etal~\cite{Goldhirsch1987},
leads to our asymptotic forms.  We find that the Lagrangian derivatives along
diverging directions of the finite-time Lyapunov exponents grow exponentially
at the characteristic rate of that direction.  This is consistent with the
intuitive notion that small displacements in those directions will be
exponentially amplified, so one expects derivatives to grow as well.  For
contracting directions, the opposite is true: the Lagrangian derivatives
converge exponentially to time-asymptotic values, but often do so at a slower
rate than the characteristic rate.  The Lagrangian derivatives of the
characteristic directions of stretching have a more complicated behaviour, and
it is not always true that a derivative along an expanding direction will
diverge---the intuitive picture fails.

The asymptotic behaviour derived using the \SVD\ method can be used to recover
so-called \emph{differential constraints} on the finite-time Lyapunov
exponents and characteristic directions.  Such constraints were first derived
in two dimensions by Tang and Boozer~\cite{Tang1996} and Giona and
Adrover~\cite{Giona1998} and were later extended to three dimensions by
Thiffeault and Boozer~\cite{Thiffeault2001}.  These earlier derivations
provided limited results and were difficult to generalise to higher
dimensions.  In particular, the convergence rate of the constraints was not
obtained, making it difficult to gauge their effect in approximations.  The
derivation given here overcomes these difficulties.

Whilst the \SVD\ method is transparent and useful for theoretical derivations
and interpretation, it is not well-suited for numerical
purposes~\cite{Geist1990}: it possesses troubling singularities and involves a
needlessly large number of equations to evolve.  A better method is the \QR\
decomposition~\cite{Goldhirsch1987,Geist1990}, also known as the
\emph{continuous Gram--Schmidt orthonormalisation} method because it is a
time-continuous version of earlier methods that involved re-orthonormalising a
set of vectors evolved using the tangent map of the
flow~\cite{Bennetin1976,Shimada1979,Eckmann1985}.  As for the \SVD\ method, we
adapt the \QR\ method to finding the Lagrangian derivatives of the finite-time
Lyapunov exponents and of the characteristic eigenvectors.  The \QR\ method
can be used to verify the differential constraints mentioned above, since they
depend on delicate cancellations that require the high accuracy afforded by
the method herein in order to be convincingly established.

The outline of the paper is as follows.  In \secref{basicth} we introduce the
basic framework and notation necessary to the subsequent development.  The
central object of study is the \emph{metric tensor} transformed to Lagrangian
coordinates, and its diagonal form that contains all the information on the
characteristic separations and directions.  \Secref{directmethod} describes
the direct method of evolving the Hessian, where its governing equations are
obtained by a variation of the ODEs and integrated directly.  This method is
not very useful numerically because the exponential blowup of the elements of
the Hessian leads to issues of limited precision, but illustrates the basic
principles and can be used to check the results of more complex methods, for
short times.

A more powerful method is introduced in \secref{SVDmethod}, the \SVD\
decomposition method.  This allows derivation of the asymptotic behaviour of
the Lagrangian derivatives.  The asymptotic behaviour is exploited in
\secref{symhesconstraints} to investigate the properties of the Hessian, and
to give a new and powerful derivation of differential constraints in chaotic
flows.

In \secref{QRmethod} the \QR\ decomposition is used to develop a suitable
numerical method for evolving the various Lagrangian derivatives.  The
numerical method is then used to verify to high precision the geometrical
constraints derived in \secref{symhesconstraints}.  \Secref{discussion}
consists of a brief summary of the main results of this paper and of
possible future work, as well as a discussion of possible applications.

\section{Characteristic Directions of Trajectory Separation}
\label{sec:basicth}

We begin with a brief overview of the concepts and notation we shall use.  We
consider the~$\sdim$-dimensional dynamical system
\begin{equation}
	\dot\xv = \vel(\xv,\time)\ ,
	\label{eq:dynsys}
\end{equation}
where the overdot indicates a time derivative, and~$\vel$ is a smooth function
of~$\xv$ and~$\time$.  The solution to~\eqref{dynsys} is a
function~$\xv(\time)$, with initial condition~$\xv(\time_0)=\lagrcv$.  We can
thus regard~$\xv(\time)$ as a coordinate transformation from the set of
initial conditions~$\lagrcv$ to the state at time~$\time$; we write this
transformation explicitly as~$\xv(\lagrcv,\time)$.\footnote{To lighten the
notation, we omit the explicit dependence of~$\xv$ on the initial
time~$\time_0$.}  Following standard terminology, we call~$\xv$ the Eulerian
coordinates and~$\lagrcv$ the Lagrangian coordinates.

The time-evolution of the Jacobian matrix~\hbox{${\Jacm^i}_p \ldef
\pd\x^i/\pd\lagrc^p$} is given by
\begin{equation}
	\dot{\Jacm^i}_p = \sum_{\ell=1}^\sdim{\dvdx^i}_\ell\,
		{\Jacm^\ell}_p,
	\label{eq:tangentode}
\end{equation}
where~${\dvdx^i}_\ell \ldef \pd\velc^i/\pd\x^\ell$; the initial conditions
are~\hbox{${\Jacm^i}_p = {\delta^i}_p$}, since the coordinates~$\xv$
and~$\lagrcv$ initially coincide.  The nonsingular Jacobian matrix tells us
how to transform vectors in~$\lagrcv$ coordinates to vectors in~$\xv$
coordinates (${\Jacm^i}_p$ is the tangent mapping to the
transformation~$\xv(\lagrcv,\time)$).  Now construct the matrix
\begin{equation}
	\metric_{pq} \ldef \sum_{\ell=1}^\sdim{\Jacm^\ell}_p\,{\Jacm^\ell}_q,
	\label{eq:metricdef}
\end{equation}
called the \emph{metric tensor in Lagrangian coordinates} or
\emph{Cauchy--Green strain tensor}~\cite{Ottino}; it is symmetric and
positive-definite, so it can be diagonalized with real positive eigenvalues
and orthogonal eigenvectors and rewritten as
\begin{equation}
	\metric_{pq} = \sum_{\sigma=1}^\sdim
		\nugr_\sigma^2\,
		(\ediruv_\sigma)_p\,(\ediruv_\sigma)_q,
	\label{eq:metricdiag}
\end{equation}
with~$\ediruv_\sigma(\lagrcv,\time)$ and~$\nugr_\sigma^2(\lagrcv,\time)$
respectively the~$\sigma$th eigenvector of~$\metric_{pq}(\lagrcv,\time)$ and
corresponding eigenvalue.  We refer to the~$\nugr_\sigma$ as the
\emph{coefficients of expansion}, because they represent the relative
deformation of the principal axes of an infinitesimal ellipsoid carried by the
flow.  The coefficients of expansion can be used to define the finite-time
Lyapunov exponents,
\begin{equation}
	\lyapexp_\sigma(\lagrcv,\time)
		\ldef \frac{1}{\time}\,\log\nugr_\sigma(\lagrcv,\time).
	\label{eq:FTLEdef}
\end{equation}
The finite-time Lyapunov exponents,~$\lyapexp_\sigma(\lagrcv,\time)$ describe
the instantaneous average rate of exponential separation of neighbouring
trajectories.\footnote{We are implicitly assuming the Euclidean norm for
vectors in Eulerian space.} The multiplicative ergodic theorem of
Oseledec~\cite{Oseledec1968} implies that the infinite-time
limit~$\lyapexpinf_\sigma$ of~$\lyapexp_\sigma(\lagrcv,\time)$ exists and is
independent of the initial condition~$\lagrcv$ in a given ergodic domain, for
almost all initial conditions.  The infinite-time
limit~$\ediruvinf_\sigma(\lagrcv)$ of the characteristic
eigenvectors~$\ediruv_\sigma(\lagrcv,\time)$ also exists but depends on the
initial condition.  The Lyapunov exponents converge very slowly, whereas for
nondegenerate exponents the characteristic eigenvectors converge exponentially
fast~\cite{Goldhirsch1987}.  The slow convergence of the Lyapunov exponents
indicates that the instantaneous separation rate of neighbouring trajectories
is not at all exponential on a typical attractor; only in the infinite-time
limit do the trajectories show a mean exponential rate of separation.  The
instantaneous deviations from this exponential rate are very large.  However,
even though it may not be growing exponentially, an eigenvalue~$\nugr_\sigma$
associated with a positive Lyapunov exponent becomes very large after a
relatively short time, and conversely an eigenvalue associated with a negative
Lyapunov exponent becomes very small.  It is thus an abuse of language, but a
convenient one we shall use, to refer to the~$\nugr_\sigma$'s as growing or
shrinking exponentially.

In this paper we shall assume that the eigenvalues~$\nugr_\sigma$ are
nondegenerate and ordered such that~\hbox{$\nugr_{\sigma-1} > \nugr_\sigma$}.
After allowing some time for chaotic behaviour to set in, we have
that~\hbox{$\nugr_\sigma \gg \nugr_\kappa$} for~$\sigma<\kappa$.  We shall
make use of this ordering often in the subsequent development.

\section{Lagrangian Derivatives: Direct Method}
\label{sec:directmethod}

When the vector field~$\vel(\xv,\time)$ is a known analytic function, explicit
evolution equations for the spatial derivatives of the~$\nugr_\mu$
and~$\ediruv_\mu$ can be derived, as pointed out in Refs.~\cite{Tang1996}
and~\cite{Tang1999b}.  The method involves expressing the derivatives
of~$\nugr_\mu$ and~$\ediruv_\mu$ in terms of derivatives of the metric
tensor~\metric.  This is done by taking the Lagrangian derivative of the
diagonal form~\eqref{metricdiag} of~\metric\ and dotting the resulting
expression with the eigenvectors~$\ediruv_\mu$.  We obtain
\begin{eqnarray}
\frac{\pd(\ediruv_\mu)_q}{\pd\lagrc^p}
	= \sum_{\sigma\ne\mu}\sum_{r,s}
	\frac{1}{\nugr_\mu^2 - \nugr_\sigma^2}\,
	(\ediruv_\mu)_r\,(\ediruv_\sigma)_s\,(\ediruv_\sigma)_q\,
	\frac{\pd\metric_{rs}}{\pd\lagrc^p}\,,\\
\frac{\pd\nugr_\mu}{\pd\lagrc^p} = \frac{1}{2\nugr_\mu}\sum_{r,s}
	(\ediruv_\mu)_r\,(\ediruv_\mu)_s\,
	\frac{\pd\metric_{rs}}{\pd\lagrc^p}\,.
\end{eqnarray}
The derivatives of the metric are obtained from the Hessian~$\Hesm^k_{qr}
\ldef \pd^2\x^k/\pd\lagrc^q\pd\lagrc^r$ of the coordinate transformation via
the relation
\begin{equation*}
	\frac{\pd\metric_{pq}}{\pd\lagrc^r}
	= \sum_{\ell=1}^\sdim\l({\Jacm^\ell}_p\,\Hesm^\ell_{qr}
	+ {\Jacm^\ell}_q\,\Hesm^\ell_{pr}\r),
\end{equation*}
obtained by differentiating the non-diagonal form~\eqref{metricdef}
of~\metric.  The Hessian is symmetric in its lower indices, and it is computed
by solving the evolution equation
\begin{equation}
	\dot\Hesm^k_{qr}
	= \sum_{\ell=1}^\sdim
		{\dvdx^k}_\ell\,
		\Hesm^\ell_{qr}
	+ \sum_{i,j=1}^\sdim\,\X^k_{i j}\,
		{\Jacm^i}_q\,
		{\Jacm^j}_r,
	\label{eq:tangent2ode}
\end{equation}
where~\hbox{$\X^k_{i j} \ldef \pd^2\velc^k/\pd\x^i\pd\x^j$}.  For the
existence and uniqueness of solutions to~\eqref{tangent2ode}, it is necessary
that~$\pd^2\velc^k/\pd\x^i\pd\x^j$ be Lipschitz, but it is sufficient
that~$\vel$ be at least thrice-differentiable.  \Eref{eq:tangent2ode} is
obtained by differentiating~\eqref{tangentode}, and the initial condition
is~$\Hesm=0$.  The time derivatives (the overdots) are taken at
constant~$\lagrcv$, so they can be commuted with Lagrangian derivatives.

The linear part of~\eqref{tangent2ode} is the same as for~\eqref{tangentode},
but now there is a nonlinear coupling term to~$\Jacm$.  Since for a chaotic
flow the matrix~$\Jacm$ has at least one exponentially growing eigenvalue, the
elements of the Hessian~$\Hesm$ ctypically grow faster than~$\Jacm$, owing to
the nonlinear coupling.  Numerically, the system~\eqref{tangent2ode} is thus
unstable in an essential way, and we can expect the integration to give
meaningless results (except along the dominant stretching direction) after the
elements of~$\Hesm$ become too large.  Nevertheless, if one is not interested
in long-time behaviour the direct method can yield satisfactory results.  It
is not suitable for a detailed, accurate, long-time solution of the Lagrangian
derivatives.  We develop such methods in \secreftwo{SVDmethod}{QRmethod}.

\section{Asymptotic Behaviour using the \SVD\ Method}
\label{sec:SVDmethod}

\subsection{Basic Method}
\label{sec:SVDbasicmethod}

Any matrix, and in particular the Jacobian matrix~$\Jacm$, can be decomposed
into the product
\begin{equation}
	\Jacm = \Usvd\Fsvd\,\transp{\Vsvd},
	\label{eq:SVDdecomp}
\end{equation}
where~$\Usvd$ and~$\Vsvd$ are orthogonal matrices and~$\Fsvd$ is diagonal.
The superscript~$\transp{{}}$ denotes a matrix transpose.  This decomposition
is called the \emph{singular value decomposition} (\SVD), and is unique up to
permutations of rows and columns.  The diagonal elements~$\nugr_\sigma$
of~$\Fsvd$ are called the singular values.  Requiring that the singular values
be ordered decreasing in size makes the decomposition unique (for
nondegenerate eigenvalues).  As can be seen by substitution
of~\eqref{SVDdecomp} into~\eqref{metricdef}, the columns of~$\Vsvd$ are
eigenvectors of~$\metric$, $\Vsvd_{q\sigma} = (\ediruv_\sigma)_q$, with
eigenvalues given by the diagonal elements of~$\transp{\Fsvd}\Fsvd$,
$(\transp{\Fsvd}\Fsvd)_{\sigma\sigma} = \nugr_\sigma^2$.  The advantage of the
\SVD\ is that it separates neatly the parts of~$\Jacm$ that are growing or
shrinking exponentially in size (as determined by the coefficients of
expansion~$\nugr_\sigma$).  The \SVD\ has the following interpretation: if we
consider an infinitesimally small ``ball'' of initial condition
obeying~\eqref{dynsys}, it will deform into an ellipsoid under the action of
the flow.  The~$\nugr_\sigma$ give the relative stretching of each principal
axis of the ellipsoid, the orthogonal matrix~$\Vsvd$ gives the principal axes
of stretching in Lagrangian coordinate space, and the orthogonal
matrix~$\Usvd$ gives the absolute orientation of the ellipse in Eulerian
space.  Constructing the metric tensor as in~\eqref{metricdef} eliminates the
Eulerian orientation, retaining the essential features of the stretching.

Greene and Kim~\cite{Greene1987} derived the equations satisfied by~$\Usvd$,
$\Vsvd$, and~$\Fsvd$:
\begin{eqnarray}
	\dot\nugr_\mu = \dvdxhat_{\mu\mu}\nugr_\mu,\label{eq:SVDnugrODE}\\
	(\transp{\Usvd}\dot\Usvd)_{\mu\nu}
	= \cases{-\frac{\dvdxhat_{\mu\nu}\nugr_\nu^2
		+ \dvdxhat_{\nu\mu}\nugr_\mu^2}{\nugr_\mu^2 - \nugr_\nu^2}
		&for $\mu \ne \nu$;\\ 0 & for $\mu = \nu$;\\}
	\label{eq:SVDUODE}\\
	(\transp{\Vsvd}\dot\Vsvd)_{\mu\nu}
	= \cases{-\frac{\nugr_\mu\,\nugr_\nu}
		{\nugr_\mu^2 - \nugr_\nu^2}\,
		\dvdxsym_{\mu\nu}
	& for $\mu \ne \nu$;\\ 0 & for $\mu = \nu$;\\}
	\label{eq:SVDVODE}
\end{eqnarray}
where
\begin{equation*}
	\dvdxhat \ldef \transp{\Usvd}\dvdx\Usvd,
	\qquad\dvdxsym \ldef \dvdxhat +	\transp{\dvdxhat}.
\end{equation*}
Numerically, the \SVD\ method has limitations: the number of quantities to
evolve is large, and when~$\nugr_\sigma \simeq \nugr_{\sigma+1}$ the
denominators become singular (See Refs.~\cite{Greene1987,Geist1990} for a
discussion of these issues).  However, conceptually the method is very
straightforward and transparent, and it gives explicit equations for all the
quantities we are interested in, as opposed to the \QR\ method
(\secref{QRmethod}) which needs to be corrected to yield the true value of
the~$\nugr_\sigma$.

For large time, when~$\nugr_\sigma/\nugr_\kappa \ll 1$,~$\sigma>\kappa$, we
can approximate~\eqref{SVDUODE} and~\eqref{SVDVODE} by
\begin{equation}
\fl
	(\transp{\Usvd}\dot\Usvd)_{\mu\nu}
	= \cases{-\dvdxhat_{\nu\mu}
		& $\mu < \nu$;\\
	+\dvdxhat_{\mu\nu}
		& $\mu > \nu$;\\
	0 & $\mu = \nu$;\\}
	\qquad\qquad
	(\transp{\Vsvd}\dot\Vsvd)_{\mu\nu}
	= \cases{-\nudec_{\mu\nu}\,\dvdxsym_{\mu\nu}
		& $\mu < \nu$;\\
	+\nudec_{\mu\nu}\,\dvdxsym_{\mu\nu}
		& $\mu > \nu$;\\
	0 & $\mu = \nu$.\\}
	\label{eq:SVDUVODEapprox}
\end{equation}
where we have used
\begin{equation}
	\nudec_{\mu\nu} \ldef
	\cases{
		{\nugr_\nu}/{\nugr_\mu} & $\mu < \nu$;\\
		{\nugr_\mu}/{\nugr_\nu} & $\mu > \nu$;\\
		\max\l(\frac{\nugr_{\mu}}{\nugr_{\mu-1}},
			\frac{\nugr_{\mu+1}}{\nugr_{\mu}}\r)
			& $\mu = \nu$.\\}
	\label{eq:nudecdef}
\end{equation}
The matrix~$\nudec$ is symmetric, and defined such that (for
nondegenerate~$\nugr$) all of its elements are decaying exponentially.
Because of the ordering of the~$\nugr_\sigma$, the diagonal
element~$\nudec_{\mu\mu}$ represents the element on the~$\mu$th row or column
that decays the slowest, that is,~$\nudec_{\mu\mu} =
\max_{\sigma\ne\mu}\nudec_{\sigma\mu}$.

If we assume that the evolution of the system takes place on some bounded
domain, then we know that~$\dvdx$, and hence~$\dvdxhat$ and~$\dvdxsym$,
remains bounded (we also assume that the vector field~$\vel$ is smooth).
Thus, the right-hand side of~$\transp{\Vsvd}\dot\Vsvd$
in~\eqref{SVDUVODEapprox} goes to zero exponentially, and we can solve the
equation perturbatively for large time,
\begin{equation*}
	\Vsvd_{q\mu} = \Vsvdinf_{q\mu} + \sum_\nu\Vsvdinf_{q\nu}
		\int_{\time_0}^{\time}
		(\transp{\Vsvd}\dot\Vsvd)_{\mu\nu}\dint\time
	+ \Order{\nudec_{\mu\mu}^2}
\end{equation*}
for some constant~$\Vsvdinf_{q\nu}$, which can only be determined by solving
the unapproximated equation~\eqref{SVDVODE}.  We conclude that for
large~$\time$ the matrix~$\Vsvd$ has the form~\cite{Goldhirsch1987}
\begin{equation}
	\Vsvd_{q\mu} = \nudec_{\mu\mu}\Vsvdt_{q\mu}
	+ \Vsvdinf_{q\mu}\,, \qquad \time\gg1.
	\label{eq:Vasym}
\end{equation}
Since~$\Vsvd_{q\mu} = (\ediruv_\mu)_q$ and~$\nudec_{\mu\mu} \rightarrow 0$,
the characteristic eigenvectors converge exponentially to their
time-asymptotic value, $(\ediruvinf_\mu)_q = \Vsvdinf_{q\mu}$.  The elements
of the matrix~$\Usvd$ do \emph{not} in general converge.  The Lyapunov
exponents have a very slow (logarithmic) convergence which does not concern us
here, as we are considering timescales of fast (roughly exponential)
convergence.

\subsection{Lagrangian Derivatives}
\label{sec:SVDLagr}

Having derived the equations of motion for the \SVD\ of~$\Jacm$, we can now
take the Lagrangian derivative of these equations of motion, in a manner
analogous to \secref{directmethod}.  We define the quantities
\begin{equation}
\fl
	\Vgrnu_{\kappa\nu} \ldef \sum_q\Vsvd_{q\kappa}\,
		\frac{\pd\log\nugr_{\nu}}{\pd\lagrc^q},
	\quad
	\VgrU_{\kappa\mu\nu} \ldef \sum_{q,i}\Vsvd_{q\kappa}\,\Usvd_{i\mu}\,
		\frac{\pd\Usvd_{i\nu}}{\pd\lagrc^q},
	\quad
	\VgrV_{\kappa\mu\nu} \ldef \sum_{q,p}\Vsvd_{q\kappa}\,\Vsvd_{p\mu}\,
		\frac{\pd\Vsvd_{p\nu}}{\pd\lagrc^q},
	\label{eq:VgrUnuVdef}
\end{equation}
which are simply the Lagrangian derivatives of~$\nugr_\nu$,~$\Usvd_{i\nu}$,
and~$\Vsvd_{p\nu}$ expressed in a convenient frame.  Note
that~\hbox{$\VgrU_{\kappa\mu\nu} = -\VgrU_{\kappa\nu\mu}$}
and~\hbox{$\VgrV_{\kappa\mu\nu} = -\VgrV_{\kappa\nu\mu}$}.  The Lagrangian
derivatives~$\VgrV_{\nu\mu\nu}$ may be regarded as the components~$\mu$ of the
curvature of the vector fields defined by the columns of~$\Vsvd_{p\nu}$.
From~\eqref{SVDnugrODE}--\eqref{SVDVODE}, we can find equations for the
evolution of~$\Vgrnu_{\kappa\nu}$,~$\VgrU_{\kappa\mu\nu}$,
and~$\VgrV_{\kappa\mu\nu}$,
\begin{eqnarray}
\fl
\dot\Vgrnu_{\kappa\nu} =
	\sum_\sigma(\transp{\Vsvd}\dot\Vsvd)_{\sigma\kappa}\,\Vgrnu_{\sigma\nu}
	+ \sum_\sigma\dvdxsym_{\sigma\nu}\,\VgrU_{\kappa\sigma\nu}
	+ \nugr_\kappa\,\Xh_{\nu\kappa\nu},
	\label{eq:VgrnuODE}\\
\fl
\dot\VgrU_{\kappa\mu\nu} =
\sum_\sigma\l[
	(\transp{\Vsvd}\dot\Vsvd)_{\sigma\kappa}\,\VgrU_{\sigma\mu\nu}
	+ (\transp{\Usvd}\dot\Usvd)_{\sigma\mu}\,\VgrU_{\kappa\sigma\nu}
	- (\transp{\Usvd}\dot\Usvd)_{\sigma\nu}\,\VgrU_{\kappa\sigma\mu}
\r]\nonumber\\
+ \sum_q\Vsvd_{q\kappa}\,\frac{\pd}{\pd\lagrc^q}
	(\transp{\Usvd}\dot\Usvd)_{\mu\nu},
	\label{eq:VgrUODE}\\
\fl
\dot\VgrV_{\kappa\mu\nu} =
\sum_\sigma\l[
	(\transp{\Vsvd}\dot\Vsvd)_{\sigma\kappa}\,\VgrV_{\sigma\mu\nu}
	+
	(\transp{\Vsvd}\dot\Vsvd)_{\sigma\mu}\,\VgrV_{\kappa\sigma\nu}
	-
	(\transp{\Vsvd}\dot\Vsvd)_{\sigma\nu}\,\VgrV_{\kappa\sigma\mu}
\r]\nonumber\\
+ \sum_q\Vsvd_{q\kappa}\,\frac{\pd}{\pd\lagrc^q}
		(\transp{\Vsvd}\dot\Vsvd)_{\mu\nu},
	\label{eq:VgrVODE}
\end{eqnarray}
where
\begin{equation}
	\Xh_{\nu\kappa\mu} \ldef \sum_{k,i,\ell}\Usvd_{k\nu}\,\Usvd_{i\kappa}
		\,\Usvd_{\ell\mu}\,\frac{\pd^2\velc^{k}}{\pd\x^i\pd\x^\ell}
\end{equation}
is symmetric in~$\kappa$ and~$\mu$.

In \apxref{SVDLagrasym}, the asymptotic behaviour of the derivatives is
obtained from the equations of motion~\eqref{VgrnuODE}--\eqref{VgrVODE}.  We
now summarise the main results of that section: for~$\time\gg 1$, by which we
mean that the dynamical system has evolved long enough for the
quantities~$\nugr_\mu$ to have reached a regime of quasi-exponential
behaviour, we have that the Lagrangian derivatives defined
by~\eqref{VgrUnuVdef} evolve asymptotically as
\begin{eqnarray}
\VgrU_{\kappa\mu\nu} = 
		\max\l(\nugr_\kappa,\nudec_{\mu\nu}\r)
		\VgrUt_{\kappa\mu\nu},
		\label{eq:VgrUasym}\\
\Vgrnu_{\kappa\nu} = 
		\max\l(\nugr_\kappa,\nudec_{\kappa\kappa},
		\nudec_{\nu\nu}\r)\Vgrnut_{\kappa\nu}
		+ \Vgrnuinf_{\kappa\nu}
	&\sim
	\max\l(\nugr_\kappa\,,1\r),
		\label{eq:Vgrnuasym}\\
\VgrV_{\kappa\mu\nu} =
		\max\l(\nudec_{\mu\nu}\nugr_\kappa,
		\nudec_{\kappa\kappa},
		\nudec_{\mu\mu},
		\nudec_{\nu\nu}\r)
		\VgrVt_{\kappa\mu\nu}
		+ \VgrVinf_{\kappa\mu\nu}
	&\sim
	\max\l(\nudec_{\mu\nu}\nugr_\kappa\,,1\r).
	\label{eq:VgrVasym}
\end{eqnarray}
Recall that in all these cases the first index,~$\kappa$, denotes the
characteristic eigendirection~$\ediruv_\kappa$ along which the Lagrangian
derivative is evaluated.  For a given~$\kappa$ with~\hbox{$\nugr_\kappa\gg
1$}, corresponding to an expanding direction of the flow,
both~$\VgrU_{\kappa\mu\nu}$ and~$\Vgrnu_{\kappa\nu}$ grow exponentially with
time (the constant~$\Vgrnuinf_{\kappa\nu}$ is then irrelevant).  Thus,
Lagrangian derivatives of~$\log\nugr_\nu$ along an expanding
direction~$\kappa$ become more singular with time, to a degree commensurate
with the separation of neighbouring initial conditions along that direction,
as given by~$\nugr_\kappa$.

Conversely, for~$\kappa$ with~\hbox{$\nugr_\kappa\ll 1$}, corresponding to a
contracting direction of the flow, both~$\VgrU_{\kappa\mu\nu}$
and~$\Vgrnu_{\kappa\nu}$ decrease exponentially with
time,~$\VgrU_{\kappa\mu\nu}$ converging toward zero and~$\Vgrnu_{\kappa\nu}$
converging to a constant,~$\Vgrnuinf_{\kappa\nu}$.  The convergence rate of
both these quantities, however, is not necessarily equal to~$\nugr_\kappa$ but
may be slower (though still converging), as denoted by the~$\max$
in~\eqref{VgrUasym} and~\eqref{Vgrnuasym}.  This slower convergence rate
of~$\Vgrnu$ has two sources: (i) From~\eqref{Vasym}, the~$\Vsvd_\kappa$ in the
definition~\eqref{VgrUnuVdef} of~$\Vgrnu_{\kappa\nu}$ converges at a
rate~$\nudec_{\kappa\kappa}$; (ii) The~$\VgrU$ term in~\eqref{VgrnuODE} limits
the convergence to~$\nudec_{\nu\nu}$.

The interpretation of the long-time behaviour of Lagrangian derivatives
of~$\Vsvd$, the characteristic eigenvectors, is less generic.  For a
contracting direction~$\kappa$, the derivatives~$\VgrV_{\kappa\mu\nu}$
converge to the constant value~$\VgrVinf_{\kappa\mu\nu}$ at a rate
of~$\max(\nudec_{\kappa\kappa}, \nudec_{\mu\mu}, \nudec_{\nu\nu})$.  This
convergence rate is dominated by the convergence of the individual~$\Vsvd$'s
in~\eqref{VgrUnuVdef}, as given by~\eqref{Vasym}. For an expanding
direction~$\kappa$, the specific behaviour of~$\VgrV_{\kappa\mu\nu}$ depends
on the relative magnitudes of the coefficients of expansion~$\nugr$.  However,
for a non-contracting direction~$\kappa$ it is always true that
\begin{equation*}
	(\VgrV_{\kappa\mu\nu} - \VgrVinf_{\kappa\mu\nu}) \ll \nugr_\kappa,
	\qquad \nugr_\kappa\ge 1,
\end{equation*}
so the gradients of~$\ediruv$ along expanding directions grow much more slowly
that those of~$\log\nugr$.

We close this section by considering the asymptotic behaviour of the
Lagrangian derivative of the determinant,~$\detmetric$, of the metric
tensor~$\metric$.  From~\eqref{metricdiag},
\begin{equation*}
	\detmetric = \sum_\nu\nugr_\nu^2\,,
\end{equation*}
so that the Lagrangian derivative of the determinant along~$\Vsvd_\kappa$ is
\begin{equation}
	\Vgrdetg_\kappa \ldef \sum_q\Vsvd_{q\kappa}\,
		\frac{\pd}{\pd\lagrc^q}\log\detmetric^{1/2}
	= \sum_\nu\Vgrnu_{\kappa\nu}\,.
\end{equation}
An equation of motion for~$\Vgrdetg_\kappa$ is obtained by
summing~\eqref{VgrnuODE} over~$\nu$, yielding
\begin{equation}
\dot\Vgrdetg_{\kappa} =
	\sum_\sigma(\transp{\Vsvd}\dot\Vsvd)_{\sigma\kappa}\,
		\Vgrdetg_{\sigma}
	+ \nugr_\kappa\,\sum_\nu\Xh_{\nu\nu\kappa},
	\label{eq:VgrdetgODE}
\end{equation}
where the~$\VgrU$ term has dropped out.  The term~$\sum_\nu\Xh_{\nu\nu\kappa}$
is proportional to~$\div\vel$, so for an incompressible flow the solution
to~\eqref{VgrdetgODE} is~\hbox{$\Vgrdetg_\kappa\equiv0$}.  But in general, for
a compressible flow with non-uniform~$\div\vel$, we find
\begin{equation}
	\Vgrdetg_{\kappa} = 
		\max\l(\nugr_\kappa,\nudec_{\kappa\kappa}\r)
			\Vgrdetgt_{\kappa}
		+ \Vgrdetginf_{\kappa},
	\qquad \time\gg 1,
	\label{eq:Vgrdetgasym}
\end{equation}
where the time dependence of~$\Vgrdetgt_\kappa$ is non-exponential.  In
contrast to~$\Vgrnu$, the limiting rate~$\nudec_{\kappa\kappa}$
in~\eqref{Vgrdetgasym} is due entirely to the convergence rate
of~$\Vsvd_\kappa$, as given by~\eqref{Vasym}.

\section{Properties of the Hessian and Constraints}
\label{sec:symhesconstraints}

\subsection{Symmetry of the Hessian}
\label{sec:symmetry-hessian}

We now make contact with the work of Dressler and Farmer~\cite{Dressler1992}
and Taylor~\cite{Taylor1993} on the form of the generalised Lyapunov
exponents, which describe the asymptotic behaviour of the Hessian.  The
Hessian, defined in \secref{directmethod}, can be recovered from the
Lagrangian derivatives of \secref{SVDLagr}.  Following Dressler and
Farmer~\cite{Dressler1992}, we project the components of the Hessian onto
the~$\Usvd$ and~$\Vsvd$ bases and define
\begin{equation}
	\Hesmh^\kappa_{\mu\nu} \ldef \sum_{\ell,p,q}
	\Usvd_{\ell\kappa}\,\Hesm^\ell_{pq}\,\Vsvd_{p\mu}\,\Vsvd_{q\nu}\,.
	\label{eq:Hesmhdef}
\end{equation}
Writing~\hbox{$\Hesm^\ell_{pq} = \pd{\Jacm^\ell}_p/\pd\lagrc^q$}, and using
the \SVD\ decomposition~\eqref{SVDdecomp} for~$\Jacm$, we find
\begin{equation}
	\Hesmh^\kappa_{\mu\nu}
	= \nugr_\kappa\,\Vgrnu_{\mu\kappa}\,\delta_{\nu\kappa}
		+ \nugr_\kappa\,\VgrV_{\mu\nu\kappa}
		+ \nugr_\nu\,\VgrU_{\mu\kappa\nu}\,.
	\label{eq:HesmSVD}
\end{equation}
The asymptotic behaviour of the Hessian is easily derived
from~\eqref{VgrUasym},~\eqref{Vgrnuasym}, and~\eqref{VgrVasym},
\begin{equation}
	\Hesmh^\kappa_{\mu\nu} = \max(\nugr_\kappa\,,\,\nugr_\mu\nugr_\nu)\,
		\Hesmt^\kappa_{\mu\nu},
	\label{eq:Hesmasym}
\end{equation}
where~$\Hesmt^\kappa_{\mu\nu}$ is a non-exponential function, as found for
maps in Refs.~\cite{Dressler1992,Taylor1993}.

Since the Hessian is symmetric in its lower indices, we could have equally
well written
\begin{equation}
	\Hesmh^\kappa_{\mu\nu}
		= \nugr_\kappa\,\Vgrnu_{\nu\kappa}\,\delta_{\mu\kappa}
		+ \nugr_\kappa\,\VgrV_{\nu\mu\kappa}
		+ \nugr_\mu\,\VgrU_{\nu\kappa\mu}\,,
	\label{eq:HesmSVDsym}
\end{equation}
where we simply interchanged~$\mu$ and~$\nu$ in~\eqref{HesmSVD}.
Equating~\eqref{HesmSVD} and~\eqref{HesmSVDsym}, we find the relations
\begin{eqnarray}
	\nugr_\mu\l(\VgrV_{\mu\mu\nu} + \Vgrnu_{\nu\mu}\r)
		= \nugr_\nu\,\VgrU_{\mu\mu\nu},
		&\mu\ne\nu,
	\label{eq:Hessymrel1}\\
	\nugr_\kappa\l(\VgrV_{\mu\nu\kappa} - \VgrV_{\nu\mu\kappa}\r)
		= \nugr_\nu\,\VgrU_{\mu\nu\kappa} -
		\nugr_\mu\,\VgrU_{\nu\mu\kappa},
		\qquad &\mu,\nu,\kappa\ \mathrm{differ}.
	\label{eq:Hessymrel2}
\end{eqnarray}
Equation~\eqref{Hessymrel1} defines~$\sdim(\sdim-1)$ independent relations,
whilst~\eqref{Hessymrel2} defines~$\sdim(\sdim-1)(\sdim-2)/2$ relations.
Thus, a total of~$\sdim^2(\sdim-1)/2$ quantities are dependent on the others
and can be eliminated, which is exactly the number of dependent components
of~$\Hesm^\kappa_{\mu\nu}$.  The relation~\eqref{Hessymrel2} can be solved
for~$\VgrV_{\kappa\mu\nu}$ to yield
\begin{equation}
\fl
	\VgrV_{\kappa\mu\nu} = \frac{1}{2}\l[
	\l(\frac{\nugr_\mu}{\nugr_\nu} + \frac{\nugr_\nu}{\nugr_\mu}\r)
		\VgrU_{\kappa\mu\nu}
	+ \l(\frac{\nugr_\kappa}{\nugr_\nu} - \frac{\nugr_\nu}{\nugr_\kappa}\r)
		\VgrU_{\mu\nu\kappa}
	+ \l(\frac{\nugr_\kappa}{\nugr_\mu} - \frac{\nugr_\mu}{\nugr_\kappa}\r)
		\VgrU_{\nu\kappa\mu}
	\r],
	\label{eq:VgrVsolved}
\end{equation}
when~$\mu$,~$\nu$, and~$\kappa$ differ.  Similarly,~\eqref{Hessymrel1} can
trivially be solved for~$\VgrV_{\mu\mu\nu}$.  Equations~\eqref{Hessymrel1}
and~\eqref{VgrVsolved} express all of the the~$\VgrV$ in terms of the~$\Vgrnu$
and~$\VgrU$.
However, the asymptotic behaviour of~$\VgrV_{\kappa\mu\nu}$ cannot be
recovered from these equations by simply substituting~\eqref{VgrUasym}
and~\eqref{Vgrnuasym}.  The reason is that the asymptotic
form~\eqref{VgrVasym} hinges on delicate cancellations between the~$\VgrU$
and~$\Vgrnu$ that are not manifest from simply looking at their equations of
motion.  For instance, in~\eqref{VgrVsolved} the coefficient of each~$\VgrU$
term grows exponentially, even though some of the~$\VgrV$'s have been shown to
converge.

Although we are not using the \SVD\ method to obtain numerical results, the
considerations of this section also apply to the \QR\ method of
\secref{QRmethod}.  The relations~\eqref{Hessymrel1} and~\eqref{Hessymrel2}
can then be used as a diagnostic tool to monitor the numerical results.

\subsection{Differential Constraints}
\label{sec:constraints}

Rather than solving for the~$\VgrV$, if the flow is chaotic the
relations~\eqref{Hessymrel1} and~\eqref{Hessymrel2} can be put to good use in
another manner.  The Lagrangian derivatives of~$\Usvd$, as contained
in~$\VgrU$, are not quantities of great interest to us.  They describe the
sensitive dependence on initial conditions of the absolute orientation of
phase-fluid elements in Eulerian space.  This information is not necessary for
solving problems in Lagrangian coordinates, and is too sensitive to initial
conditions to be of use anyhow.  We thus substitute the time-asymptotic form
of~$\VgrU$, given by~\eqref{VgrUasym}, in the right-hand side
of~\eqref{Hessymrel1} and~\eqref{Hessymrel2}, yielding
\begin{eqnarray}
	\VgrV_{\mu\mu\nu} + \Vgrnu_{\nu\mu}
		= \max\l(\nugr_\nu\,,\frac{\nugr_\nu}{\nugr_\mu}
		\,\,\nudec_{\mu\nu}\r)\VgrUt_{\mu\mu\nu}\,,
	\label{eq:aconstrtype1}\\
	\VgrV_{\mu\nu\kappa} - \VgrV_{\nu\mu\kappa}
		= \frac{\nugr_\nu}{\nugr_\kappa}\,
			\max\l(\nugr_\mu\,,\nudec_{\nu\kappa}\r)
			\VgrUt_{\mu\nu\kappa}
		- \frac{\nugr_\mu}{\nugr_\kappa}\,
			\max\l(\nugr_\nu\,,\nudec_{\mu\kappa}\r)
			\VgrUt_{\nu\mu\kappa}\,,
	\label{eq:aconstrtype2}
\end{eqnarray}
where, as before,~$\mu$, $\nu$, and~$\kappa$ differ.  Below we show that under
certain conditions the right-hand side of~\eqref{aconstrtype1}
or~\eqref{aconstrtype2} converges toward zero, giving us asymptotic
\emph{differential constraints} on~$\Vgrnu$ and~$\VgrV$.

\subsubsection{Type I Constraints}
\label{sec:type1constr}

For~$\mu<\nu$,~\eqref{aconstrtype1} can be written
\begin{equation}
	\VgrV_{\mu\mu\nu} + \Vgrnu_{\nu\mu}
		= \max\l(\nugr_\nu\,,\,\nudec_{\mu\nu}^2\r)
		\VgrUt_{\mu\mu\nu}\,,
	\qquad \mu<\nu,
	\label{eq:constrtype1}
\end{equation}
If the index~$\nu$ corresponds to a contracting direction ($\nugr_\nu\ll1$),
the right-hand side of~\eqref{constrtype1} goes to zero exponentially fast, at
a rate~$\nugr_\nu$ or~$\nudec_{\mu\nu}^2$, whichever is slowest.  In that
case~\eqref{constrtype1} is a \emph{constraint} implying that for
large~$\time$ we have
\begin{equation}
	(\VgrV_{\mu\mu\nu} + \Vgrnu_{\nu\mu}) \longrightarrow 0,
	\qquad \nugr_\nu\ll\nugr_\mu\,,\ \ \nugr_\nu\ll 1.
	\label{eq:constrtype1b}
\end{equation}
We refer to~\eqref{constrtype1b} as type I constraints.  The total number of
such constraints is
\begin{equation}
	\NI = \ncontr\l[\sdim-\half(\ncontr+1)\r],
	\label{eq:NI}
\end{equation}
where~$\sdim$ is the dimension of the space and~$\ncontr$ is the number of
contracting directions (i.e., the number of negative Lyapunov exponents)
possessed by the flow in a particular ergodic domain.  \Tabref{NI}
gives the number of type I constraints,~$\NI$, as a function of~$\sdim$
and~$\ncontr$.
\begin{table}
\caption{The total number of type I constraints for low-dimensional systems,
as given by~\eqref{NI}.  The rows denote~$\sdim$, the columns the number of
contracting directions~$\ncontr\le\sdim$.}
\label{table:NI}
\vspace{1em}
\begin{center}
\begin{tabular}{ccccccccc}
\hline
 && \textsl{1} & \textsl{2} & \textsl{3} & \textsl{4} & \textsl{5} &
\textsl{6} & \textsl{7}\\
\hline
\textsl{1} &\quad& 0 & & & & & &\\
\textsl{2} && 1 & 1 &  & & & &\\
\textsl{3} && 2 & 3 & 3 & & & &\\
\textsl{4} && 3 & 5 & 6 & 6 & & &\\
\textsl{5} && 4 & 7 & 9 & 10 & 10 & &\\
\textsl{6} && 5 & 9 & 12 & 14 & 15 & 15 &\\
\textsl{7} && 6 & 11 & 15 & 18 & 20 & 21 & 21\\
\hline
\end{tabular}
\end{center}
\end{table}

In two dimensions, we typically have one contracting direction, so there is a
single type I constraint.  This is the same constraint that was derived in
Refs.~\cite{Tang1996,Thiffeault2001}.

In three dimensions, for an autonomous flow, we typically also have one
contracting direction.  There are then two type I constraint.  These
constraints correspond to those derived in Ref.~\cite{Thiffeault2001}.

A special case of the type I constraints is obtained by
setting~\hbox{$\nu=\sdim$} in~\eqref{constrtype1}, and then summing
over~$\mu<\sdim$, to yield
\begin{equation}
\fl
	\sum_{q}\frac{1}{\detmetric^{1/2}}\,
	\frac{\pd}{\pd\lagrc^q}\l(\detmetric^{1/2}\,(\ediruv_{\sdim})_q\r)
	- \sum_q(\ediruv_{\sdim})_q\frac{\pd}{\pd\lagrc^q}\log\nugr_\sdim
	\sim \max\l(\nugr_\sdim\,,\,\nudec_{\sdim\sdim}^2\r)
	\rightarrow 0.
	\label{eq:divsconstr}
\end{equation}
This constraint was discovered numerically and used by Tang and
Boozer~\cite{Tang1996,Tang1999b,Tang1999a,Tang2000} and derived in three
dimensions by Thiffeault and Boozer~\cite{Thiffeault2001}.  It has been used
to study the anticorrelation between curvature and
stretching~\cite{Tang1996,Tang1999b,Thiffeault2001} and to transform the
advection--diffusion equation in Lagrangian coordinates to an approximate
one-dimensional form~\cite{Thiffeault2001f}.  (See also \apxref{Eulerian} for
an Eulerian version of the constraint.)  The present method not only gives the
constraint in a very direct manner for any dimension~$\sdim$, but it also
provides us with its asymptotic convergence rate, as determined by the
right-hand side of~\eqref{divsconstr}.  It also shows that the
constraint~\eqref{divsconstr} does not stand alone, but is the sum of several
independent constraints.  More details on the differences between this paper
and earlier approaches are given in \secref{curvature}.

\subsubsection{Type II Constraints}
\label{sec:type2constr}

Equation~\eqref{aconstrtype2} implies that
\begin{equation}
	\VgrV_{\mu\nu\kappa} - \VgrV_{\nu\mu\kappa}
		\sim
		\max\l(\frac{\nugr_\mu\,\nugr_\nu}{\nugr_\kappa}\,,
		\frac{\nugr_\nu}{\nugr_\kappa}\,\nudec_{\nu\kappa}\,,
		\frac{\nugr_\mu}{\nugr_\kappa}\,\nudec_{\mu\kappa}\r).
	\label{eq:aaconstrtype2}
\end{equation}
We are interested in finding constraints analogous to the type I constraints
of~\secref{type1constr}.  It is clear that unless both~$\mu$ and~$\nu$ are
greater than~$\kappa$, the right-hand side of~\eqref{aaconstrtype2} is of
order unity or greater, and so does not go to zero.  We can assume without
loss of generality that~$\mu<\nu$, so that
\begin{equation}
	\VgrV_{\mu\nu\kappa} - \VgrV_{\nu\mu\kappa} \sim
		\nudec_{\mu\kappa}\,\max\l(\nugr_\nu\,,
		\nudec_{\mu\kappa}\r),\qquad
	\kappa<\mu<\nu,
	\label{eq:constrtype2}
\end{equation}
where we have used~$\nudec_{\nu\kappa}\ll\nudec_{\mu\kappa}$.  Whether or
not~\eqref{constrtype2} is a constraint depends on the specific behaviour
of~$\nugr_\mu\,\nugr_\nu/\nugr_\kappa$.  Clearly, we have a constraint
if~\hbox{$\nugr_\nu\ll 1$}, since~\hbox{$\nudec_{\mu\kappa}\ll 1$}.  This
provides a lower bound on the number~$\NII$ of type II constraints; by
choosing~$\nu$ from the~$\ncontr$ contracting directions, and summing over the
remaining~\hbox{$\kappa<\mu<\nu$}, we obtain
\begin{equation}
	\NII \ge
	\half\ncontr\l[\sdim^2 - (\ncontr + 2)
		\l(\sdim - \third(\ncontr + 1)\r)\r].
	\label{eq:NIImin}
\end{equation}
But even if~\hbox{$\nugr_\nu\gg1$} we can have a constraint, as long
as~\hbox{$\nugr_\nu\,\nudec_{\mu\kappa}\ll1$}.  This depends on the particular
problem at hand; hence,~\eqref{NIImin} is only a lower bound, but a fairly
tight one for low dimensions.  \Tabref{NIImin} enumerates the minimum number
of type II constraints as a function of~$\sdim$ and~$\ncontr$.
\begin{table}
\caption{Lower bound on the number of type II constraints, as given
by~\eqref{NIImin}.  The rows denote~$\sdim$, the columns the number of
contracting directions~$\ncontr\le\sdim$.}
\label{table:NIImin}
\vspace{1em}
\begin{center}
\begin{tabular}{ccccccccc}
\hline
 && \textsl{1} & \textsl{2} & \textsl{3} & \textsl{4} & \textsl{5} &
\textsl{6} & \textsl{7}\\
\hline
\textsl{1} &\quad& 0 & & & & & &\\
\textsl{2} && 0 & 0 &  & & & &\\
\textsl{3} && 1 & 1 & 1 & & & &\\
\textsl{4} && 3 & 4 & 4 & 4 & & &\\
\textsl{5} && 6 & 9 & 10 & 10 & 10 & &\\
\textsl{6} && 10 & 16 & 19 & 20 & 20 & 20 &\\
\textsl{7} && 15 & 25 & 31 & 34 & 35 & 35 & 35\\
\hline
\end{tabular}
\end{center}
\end{table}

Note that when~$\kappa$, $\mu$, and $\nu$ differ we can write
\begin{equation}
	\VgrV_{\mu\nu\kappa} - \VgrV_{\nu\mu\kappa} =
		-\sum_q(\ediruv_\kappa)_q\,\l[\ediruv_\mu\,,\ediruv_\nu\r]_q,
\end{equation}
where the Lie bracket is
\begin{equation}
	\l[\ediruv_\mu\,,\ediruv_\nu\r]_q \ldef
	\sum_p(\ediruv_\mu)_p\,\frac{\pd}{\pd\lagrc^p}\,(\ediruv_\nu)_q
	- \sum_p(\ediruv_\nu)_p\,\frac{\pd}{\pd\lagrc^p}\,(\ediruv_\mu)_q\,.
	\label{eq:Liebrakdef}
\end{equation}
The type II constraints are thus forcing certain Lie brackets of the
characteristic directions~$\ediruv_\sigma$ to vanish asymptotically.  The
geometrical implications of this, and perhaps a connexion to the Frobenius
theorem~\cite{Schutz} and the existence of submanifolds, remains to be
explored.

If we restrict to three dimensions, then there is at least one type II
constraint; it can be written as
\begin{equation}
	\VgrV_{231} - \VgrV_{321} =
		\ediruv_1\cdot\curllc\ediruv_1,
	\label{eq:constrtype2-3D}
\end{equation}
where~$\gradlc$ denotes a gradient with respect to the Lagrangian
coordinates~$\lagrcv$.  This constraint is the special case that was derived
in Ref.~\cite{Thiffeault2001}.

\subsection{Riemannian Curvature}
\label{sec:curvature}

We now compare the approach of \secref{constraints} to the earlier attempts of
Refs.~\cite{Tang1996,Thiffeault2001}, where the constraints were derived for
two and three dimensional flows by examining the form of the Riemann curvature
tensor associated with the metric~$\metric$.  We list the advantages of the
present method.

Nontrivial metrics can have curvature; a straightforward method of computing
that tensor is through the use of \emph{Ricci rotation
coefficients}~\cite{Wald},
\begin{equation}
	\Riccirotc_{\kappa\mu\nu} \ldef
	\sum_{i,j}\Usvd_{i\kappa}\,\Usvd_{j\mu}\,
	\frac{\pd}{\pd\x^i}\,\Usvd_{j\nu}\,.
\end{equation}
These satisfy the antisymmetry property~\hbox{$\Riccirotc_{\kappa\mu\nu} =
-\Riccirotc_{\kappa\nu\mu}$}, and can be rewritten in terms of the~$\VgrU$
of~\eqref{VgrUnuVdef} as
\begin{equation}
	\Riccirotc_{\kappa\mu\nu} = \nugr_\kappa^{-1}\,\VgrU_{\kappa\mu\nu}\,.
	\label{eq:rotcVgrU}
\end{equation}
In terms of the rotation coefficients, the Riemann curvature tensor
is~\cite[p.~51]{Wald}
\begin{eqnarray}
\fl
	\Rcurv_{\mu\nu\kappa\sigma} =
	\sum_i\Usvd_{i\kappa}\,\frac{\pd}{\pd\x^i}\,
		\Riccirotc_{\sigma\mu\nu}
	- \sum_i\Usvd_{i\sigma}\,\frac{\pd}{\pd\x^i}\,
		\Riccirotc_{\kappa\mu\nu}\nonumber\\
	- \sum_{\tau}\l[
	\Riccirotc_{\kappa\tau\mu}\,\Riccirotc_{\sigma\tau\nu}
	- \Riccirotc_{\sigma\tau\mu}\,\Riccirotc_{\kappa\tau\nu}
	+ \Riccirotc_{\kappa\tau\sigma}\,\Riccirotc_{\tau\mu\nu}
	- \Riccirotc_{\sigma\tau\kappa}\,\Riccirotc_{\tau\mu\nu}
	\r].
	\label{eq:Rcurvrotc}
\end{eqnarray}
If we use the relations~\eqref{Hessymrel1} and~\eqref{Hessymrel2} to solve
for~$\VgrU$ in terms of~$\VgrV$ and~$\Vgrnu$, we can rewrite~\eqref{rotcVgrU}
as
\begin{eqnarray}
\fl
	\Riccirotc_{\kappa\mu\nu} =
	\frac{1}{2\nugr_\kappa\nugr_\mu\nugr_\nu}
	\l\{\nugr_\mu^2(\VgrV_{\nu\kappa\mu} - \VgrV_{\kappa\nu\mu})
	+ \nugr_\nu^2(\VgrV_{\kappa\mu\nu} - \VgrV_{\mu\kappa\nu})
	- \nugr_\kappa^2(\VgrV_{\mu\nu\kappa} - \VgrV_{\nu\mu\kappa})\r\}
	\nonumber\\*
	+ \frac{1}{\nugr_\nu}\,\delta_{\mu\kappa}\,\Vgrnu_{\nu\mu}
	- \frac{1}{\nugr_\mu}\,\delta_{\nu\kappa}\,\Vgrnu_{\mu\nu}\,.
	\label{eq:rotcVgrVnu}
\end{eqnarray}
The form of the curvature obtained by inserting~\eqref{rotcVgrVnu} into
\eqref{Rcurvrotc} is essentially the one obtained in three dimensions in
Ref.~\cite{Thiffeault2001}.  In Ref.~\cite{Tang1996}, the curvature was
calculated directly from the Christoffel symbols.  The constraints were then
deduced by imposing the boundedness of the curvature tensor: some terms
in~\eqref{rotcVgrVnu} would appear to grow exponentially, so their coefficient
must go to zero to maintain a finite curvature.  In this manner, the type I
and type II constraints were derived in two and three dimensions, backed by
numerical evidence~\cite{Tang1996,Thiffeault2001}.

The approach used in the present paper to derive the constraints is
advantageous in several ways: (i) It is valid in any number of dimensions;
(ii) It avoids using the curvature, which is difficult to compute; (iii) There
are no assumptions about the growth rate of individual terms in the
curvature~\cite{Thiffeault2001}; (iv) The convergence rates of the constraints
are given explicitly [\eqref{constrtype1} and~\eqref{constrtype2}]; (v) The
number of constraints can be predicted [\eqref{NI} and~\eqref{NIImin}].  The
crux of the difference between the two approaches is that here we use the
variable~$\VgrU$ to estimate the asymptotic behaviour of the constraints
directly, rather than relying on indirect evidence from the curvature.  Thus,
the derivation of the time-asymptotic form of~$\VgrU$ is essential.

\section{Numerical Computations using the \QR\ Method}
\label{sec:QRmethod}

\subsection{Basic Method}
\label{sec:QRbasicmethod}

The \QR\ method, like the \SVD\ method, avoids the numerical problems
associated with evolving the Jacobian matrix~$\Jacm$ by using a judicious
matrix decomposition.  The \QR\ decomposition says that any matrix, and in
particular~$\Jacm$, can be written
\begin{equation}
	\Jacm = \Qqr\Rqr,
	\label{eq:QRdecomp}
\end{equation}
where~$\Qqr$ is an orthogonal matrix and~$\Rqr$ is upper-triangular.  For our
case,~$\Rqr$ has positive diagonal elements.  The \QR\ decomposition method of
finding Lyapunov exponents is also called the continuous Gram--Schmidt
orthonormalisation method by some
authors~\cite{Goldhirsch1987,Christiansen1997}, referring to the matrix~$\Qqr$
being obtained from~$\Jacm$ by the Gram--Schmidt method.  The \QR\ method is
an approximate version of the \SVD\ method.  The matrix~$\Qqr$ is analogous
to~$\Usvd$ in that it embodies the Eulerian information about the orientation
of the ellipsoid (see \secref{SVDbasicmethod}), and drops out of~$\metric$ as
required.  But the resulting expression~$\metric=\transp{\Rqr}\Rqr$ does not
manifestly give a diagonalisation of~$\metric$.  Below in~\eqref{GSorthoqr}
and~\eqref{qreigensystem} we give the eigenvectors~$\ediruv_\sigma$ and
coefficients of expansion~$\nugr_\sigma$ in terms of~$\Rqr$, though the
expression is not exact but is exponentially accurate with time (see
\apxref{QReigensystem}).

Let
\begin{equation}
	\nuqr_\mu \ldef \Rqr_{\mu\mu}, \qquad
	\Dqr_{\mu\nu} \ldef \nuqr_\mu\,\delta_{\mu\nu}, \qquad
	\Rqr_{\mu q} \rdef \Dqr_{\mu\mu}\,\rqr_{\mu q}.
\end{equation}
That is,~$\nuqr$ is a vector containing the diagonal elements of~$\Rqr$,
$\Dqr$ is a diagonal matrix with the~$\nuqr$ along the diagonal, and~$\rqr$
is~$\Rqr$ with the~$\mu$th row rescaled by~$\nuqr_\mu$.  The time-evolution of
these quantities is~\cite{Goldhirsch1987,Geist1990}
\begin{eqnarray}
\bs
\dot\nuqr_\mu = \Gqr_{\mu\mu}\nuqr_\mu;
	\label{eq:nuqrODE}\\
(\transp{\Qqr}\dot\Qqr)_{\mu\nu}
	= \cases{-\Gqr_{\nu\mu}	& $\mu < \nu$;\\
	+\Gqr_{\mu\nu} & $\mu > \nu$;\\
	0 & $\mu = \nu$;\\}
	\label{eq:QqrODE}\\
\dot\rqr_{\mu q} = \sum_{\sigma=\mu+1}^q\,\frac{\nuqr_\sigma}{\nuqr_\mu}\,
		\Aqr_{\mu\sigma}\,\rqr_{\sigma q},
	\qquad \mu < q;
	\label{eq:rqrODE}
\end{eqnarray}
where
\begin{equation}
	\Gqr \ldef \transp{\Qqr}\dvdx\Qqr,
	\qquad\Aqr \ldef \Gqr + \transp{\Gqr}.
\end{equation}
Equation~\eqref{nuqrODE} is identical to~\eqref{SVDnugrODE} for~$\nugr_\mu$ in
the \SVD\ method, and~\eqref{QqrODE} is identical to the time-asymptotic form
of~\eqref{SVDUODE} for~$\Usvd$, given by~\eqref{SVDUVODEapprox}.  Hence, we
expect~$\nuqr_\mu$ and~$\nugr_\mu$ to have similar asymptotic behaviour,
though their exact value differs.

Unlike the \SVD\ method, in the \QR\ method the eigenvalues~$\nugr_\mu$ and
eigenvectors~$\ediruv_{\mu}$ are not evolved directly.  They can be recovered
from the~$\nuqr_\mu$ and the matrix~$\rqr$ in the following manner.
Let~$\GSqr$ be the lower-triangular matrix that effects the Gram--Schmidt
orthonormalisation of~$\rqr$, that is
\begin{equation}
	\Wqr_{q\mu} = \sum_{\tau=1}^\mu\GSqr_{\mu\tau}\,\rqr_{\tau q}\,,
	\label{eq:GSorthoqr}
\end{equation}
where~$\GSqr_{\mu\nu} = 0$ for $\mu<\nu$, and~$\Wqr$ is orthogonal.  The
eigenvectors of~$\metric$ and corresponding coefficients of expansion are then
\begin{equation}
	(\ediruv_{\mu})_q = \Wqr_{q\mu}\,, \qquad
	\nugr_\mu = \frac{\nuqr_\mu}{\GSqr_{\mu\mu}}\,,
	\label{eq:qreigensystem}
\end{equation}
to exponential accuracy with time (the relative error on~$\ediruv_\mu$
and~$\nugr_\mu^2$ is of order~$(\nuqr_\mu/\nuqr_{\mu-1})^2$).
Equation~\eqref{qreigensystem} is proved in \apxref{QReigensystem}.

By definition, the matrix~$\Wqr$ is obtained by Gram--Schmidt
orthonormalisation of the upper-triangular matrix~$\rqr$.  In performing this
orthonormalisation, we have to compute the diagonal elements~$\GSqr_{\mu\mu}$,
so there is no extra work involved in correcting the~$\nuqr_\mu$ if we are
calculating the eigenvectors~$\Wqr$.  Note that this Gram--Schmidt procedure
does not represent an extra overhead in solving the system of
ODEs~\eqref{nuqrODE}--\eqref{rqrODE}, as the orthonormalisation need only be
effected at the end of the integration, when the eigenvectors are required.
This orthonormalisation should not be confused with the continuous
Gram--Schmidt orthonormalisation of the \QR\ method, whose purpose is to
evolve the orthogonal frame given by~$\Qqr$.

In most examples of applications of the \QR\ method, the correction derived
above to the eigenvalues is omitted~\cite{Greene1987,Geist1990}.  The reason
for this is that typically what is sought are the infinite-time Lyapunov
exponents,
\begin{equation*}
	\lyapexpinf_\mu = \lim_{\time\rightarrow\infty}\frac{1}{\time}
		\l(\log\nuqr_\mu - \log\GSqr_{\mu\mu}\r).
\end{equation*}
Since the~$\GSqr_{\mu\mu}$ converge to constant values, they are irrelevant to
the asymptotic value~$\lyapexpinf_\mu$.  This means that it is possible to
find the infinite-time Lyapunov exponents without solving for the
eigenvectors.  However, as mentioned in \secref{basicth}, we are interested
here in timescales much shorter than the convergence time of Lyapunov
exponents, so we include the correction.

We close this section by giving an explicit recurrence relation for
the~$\Wqr_{q\mu}$ and the lower-triangular matrix~$\GSqr$:
\begin{eqnarray}
	\Wqr_{p\mu} = \GSqr_{\mu\mu}\biggl[
		\rqr_{\mu p} - \sum_{\sigma=1}^{\mu-1}
		\sum_q \rqr_{\mu q}\Wqr_{q\sigma}\,\Wqr_{p\sigma}
	\biggr],
	\label{eq:WGSdef}\\
	\GSqr_{\mu\mu} = \biggl[\sum_{q}\rqr_{\mu q}^2
		- \sum_{\sigma=1}^{\mu-1}
		\Bigl(\sum_{q}\rqr_{\mu q}\Wqr_{p\sigma}\Bigr)^2\biggr]^{-1/2},
	\label{eq:adiagGSdef}\\
	\GSqr_{\mu\nu} = -\GSqr_{\mu\mu}\sum_{\sigma=\nu}^{\mu-1}\sum_q
		\rqr_{\mu q}\Wqr_{q \sigma}\,\GSqr_{\sigma\nu},
		\qquad\mu>\nu.
	\label{eq:aGSdef}
\end{eqnarray}
These follow from the usual Gram--Schmidt procedure.


\subsection{Lagrangian Derivatives}
\label{sec:QRLagr}

We now proceed to obtain ordinary differential equations for the derivatives
of the eigenvalues and eigenvectors of~$\metric$, using the \QR\ method.
Define
\begin{equation*}
\fl
	\Wgrnu_{\kappa\nu} \ldef \sum_q\Wqr_{q\kappa}\,
		\frac{\pd\log\nuqr_\nu}{\pd\lagrc^q},
	\qquad
	\WgrQ_{\kappa\mu\nu} \ldef \sum_q\Wqr_{q\kappa}\,\Qqr_{k\mu}
		\,\frac{\pd\Qqr_{k\nu}}{\pd\lagrc^q},
	\qquad
	\Wgrr_{\kappa\mu p} \ldef \sum_q\Wqr_{q\kappa}
		\,\frac{\pd\rqr_{\mu p}}{\pd\lagrc^q},
\end{equation*}
where~$\WgrQ_{\kappa\mu\nu} = -\WgrQ_{\kappa\nu\mu}$.  The tensors~$\Wgrnu$
and~$\WgrQ$ are the \QR\ method analogues of~$\Vgrnu$ and~$\VgrU$ defined
in~\eqref{VgrUnuVdef} for the \SVD\ method.  The tensor~$\Wgrr$ has no
analogue in the \SVD\ method, but is used to obtain the Lagrangian derivatives
of the eigenvectors~$\Wqr_{q\mu}$.

Using the equations of motion~\eqref{nuqrODE}--\eqref{rqrODE}
for~$\nuqr$,~$\Qqr$, and~$\rqr$, we find
\begin{eqnarray}
\fl
\dot\Wgrnu_{\kappa\nu} = \sum_\sigma(\transp{\Wqr}\dot\Wqr)_{\sigma\kappa}
		\,\Wgrnu_{\sigma\nu}
		+ \sum_\sigma\Aqr_{\sigma\nu}\WgrQ_{\kappa\sigma\nu}
		+ \Yt_{\kappa\nu\nu}
	\label{eq:WgrnuODE}\\
\fl
\dot\WgrQ_{\kappa\mu\nu} =
	\sum_\sigma(\transp{\Wqr}\dot\Wqr)_{\sigma\kappa}
		\,\WgrQ_{\sigma\mu\nu}
	+ (\Gqr_{\nu\nu} - \Gqr_{\mu\mu})\WgrQ_{\kappa\mu\nu}\nonumber\\
	- \sum_{\sigma<\mu}\Aqr_{\sigma\mu}\,\WgrQ_{\kappa\sigma\nu}
	+ \sum_{\sigma>\nu}\Aqr_{\sigma\nu}\,\WgrQ_{\kappa\mu\sigma}
	- \Yt_{\kappa\nu\mu}\,,
	\quad &\mu<\nu,
	\label{eq:WgrQODE}\\
\fl
\dot\Wgrr_{\kappa\mu q} = \sum_\sigma(\transp{\Wqr}\dot\Wqr)_{\sigma\kappa}
		\,\Wgrr_{\sigma\mu q}\nonumber\\
+ \sum_{\sigma=\mu+1}^q \frac{\nuqr_\sigma}{\nuqr_\mu}
	\Bigl[(\Wgrnu_{\kappa\sigma} - \Wgrnu_{\kappa\mu})\,
		\rqr_{\sigma q}\Aqr_{\sigma\mu}
	+ \Aqr_{\sigma\mu}\,\Wgrr_{\kappa\sigma q}\nonumber\\
\qquad\qquad + \rqr_{\sigma q}\l(\Aqr_{\tau\mu}\,\WgrQ_{\kappa\tau\sigma}
	+ \Aqr_{\tau\sigma}\,\WgrQ_{\kappa\tau\mu}
	+ \Yt_{\kappa\sigma\mu} + \Yt_{\kappa\mu\sigma}\r)\Bigr]\,,
	\qquad &\mu < q.
	\label{eq:WgrrODE}
\end{eqnarray}
The driving term~$\Yt$ is
\begin{equation*}
	\Yt_{\kappa\mu\nu} \ldef \sum_{\sigma=\kappa}^\sdim
	(\GSqr^{-1})_{\sigma\kappa}\,\nuqr_\sigma\,\Xt_{\mu\nu\sigma}\,,
\end{equation*}
where
\begin{equation*}
	\Xt_{\nu\kappa\mu} \ldef \sum_{k,i,\ell}\Qqr_{k\nu}\,\Qqr_{i\kappa}
		\,\Qqr_{\ell\mu}\,\frac{\pd^2\velc^{k}}{\pd\x^i\pd\x^\ell}
\end{equation*}
is analogous to~$\Xh$ of the \SVD\ method, and is also symmetric in~$\kappa$
and~$\mu$.  The lower-triangular matrix~$\GSqr^{-1} = \rqr\,\Wqr$ was defined
in~\eqref{GSorthoqr}.

In order to solve~\eqref{WgrnuODE}--\eqref{WgrrODE}, we need to obtain the
derivatives~$\transp{\Wqr}\dot\Wqr$.  Because~$\Wqr$ is obtained from~$\rqr$
via Gram--Schmidt orthonormalisation, the time derivatives of~$\Wqr$ are
deduced from those of~$\rqr$ by differentiation of~\eqref{WGSdef}.  After
multiplying that equation by~$\Wqr_{p\nu}$, with $\mu < \nu$, we find
\begin{equation}
\fl
	(\transp{\Wqr}\dot\Wqr)_{\mu\nu}
	= -\GSqr_{\mu\mu}\l[\,\sum_{p=\mu+1}^\sdim
	\Wqr_{p\nu}\,\dot\rqr_{\mu p}
	+ \sum_{\sigma = 1}^{\mu-1} (\GSqr^{-1})_{\mu\sigma}
		(\transp{\Wqr}\dot\Wqr)_{\sigma\nu}
	\r], \qquad \mu < \nu.
	\label{eq:WdotWGS}
\end{equation}
Owing to the orthogonality of~$\Wqr$, the
matrix~$(\transp{\Wqr}\dot\Wqr)_{\mu\nu}$ is antisymmetric.
Equation~\eqref{WdotWGS} defines a recurrence relation for
the~$(\transp{\Wqr}\dot\Wqr)_{\mu\nu}$, starting
with~$(\transp{\Wqr}\dot\Wqr)_{1\nu}$, in terms of the time derivatives
of~$\rqr$, given by~\eqref{rqrODE}.

The recipe for finding the Lagrangian derivatives thus consists of
solving~\eqref{dynsys},~\eqref{nuqrODE}--\eqref{rqrODE}
and~\eqref{WgrnuODE}--\eqref{WgrrODE} using a standard ODE integration
scheme. In doing so, use must be made of the Gram--Schmidt
procedure~\eqref{WGSdef}, which yields~$\Wqr$ and consequently also~$\GSqr$
via~\hbox{$\GSqr^{-1} = \rqr\,\Wqr$}.  The matrix~$\GSqr$ must then be
inserted into the recurrence relation~\eqref{WdotWGS}
for~$\transp{\Wqr}\dot\Wqr$, allowing finally the full evaluation of the
right-hand side of~\eqref{WgrnuODE}--\eqref{WgrrODE}.  The total number of
ODEs involved is~\hbox{$\sdim(2\sdim^2 + 3\sdim + 3)/2$}; in two dimensions,
this is 17, in three, 45.  In evaluating the right-hand side
of~\eqref{WgrnuODE}--\eqref{WgrrODE}, the most expensive term to evaluate
is~$\Yt$, which scales as~$\sdim^4$, obfuscating the cost of the Gram--Schmidt
procedures for~$\Wqr$ and~$\transp{\Wqr}\dot\Wqr$.  It is thus clear that this
numerical method is not well suited to higher-dimensional dynamical systems.
However, it is appropriate to applications such as chaotic mixing,
where~$\vel$ is a two- or three-dimensional flow.

We are not quite done yet: even though we can now solve the ODEs, they do not
give the Lagrangian derivatives of the~$\nugr_\mu$ and~$\Wqr_{q\mu}$ directly.
The~$\Wqr_{q\mu}$ are obtained from the~$\rqr_{\nu p}$ via Gram--Schmidt
orthonormalisation, so we need to proceed as we did for the time derivatives
of~$\Wqr$,~\eqref{WdotWGS}, and take a Lagrangian derivative
of~\eqref{WGSdef}.  We obtain the recurrence relation
\begin{equation}
	\WgrW_{\kappa\mu\nu}
	= -\GSqr_{\mu\mu}\l[\,\sum_{p=\mu+1}^\sdim
	\Wqr_{p\nu}\,\Wgrr_{\kappa\mu p}
	+ \sum_{\sigma = 1}^{\mu-1} (\GSqr^{-1})_{\mu\sigma}
		\WgrW_{\kappa\sigma\nu}
	\r], \qquad \mu < \nu,
	\label{eq:WgrWGS}
\end{equation}
where
\begin{equation*}
	\WgrW_{\kappa\mu\nu} \ldef \sum_{p,q}\Wqr_{q\kappa}\,\Wqr_{p\mu}
		\,\frac{\pd\Wqr_{p\nu}}{\pd\lagrc^q}
\end{equation*}
is the analogue to~$\VgrV$ in the \SVD\ method.  The recurrence relation is
solved by first evaluating~$\WgrW_{\kappa 1\nu}$ and then incrementing~$\mu$,
always keeping~$\mu<\nu$.  The antisymmetry of~$\WgrW_{\kappa\mu\nu}$ in~$\mu$
and~$\nu$ means that we do not need to consider the~$\mu>\nu$ case.

Finally, we need the Lagrangian derivative of~$\GSqr_{\mu\mu}$ in order to
find the derivative of~$\nugr_\mu$.  Indeed, because of the correction
to~$\nuqr_\mu$ given in~\eqref{qreigensystem}, we have
\begin{equation*}
	\Vgrnu_{\kappa\nu} = \Wgrnu_{\kappa\nu} - \Wgra_{\kappa\nu}\,,
\end{equation*}
where~$\Vgrnu_{\kappa\nu}$ was defined in~\eqref{VgrUnuVdef}, and
\begin{equation*}
	\Wgra_{\kappa\nu} \ldef \sum_q\Wqr_{q\kappa}\,
		\frac{\pd\log\GSqr_{\nu\nu}}{\pd\lagrc^q}
\end{equation*}
is the correction.  The explicit form for~$\Wgra$ is readily obtained in the
same manner as~\eqref{WgrWGS} by differentiating~\eqref{adiagGSdef}, to yield
\begin{equation}
	\Wgra_{\kappa\nu}
	= -\GSqr_{\nu\nu}\l[\,\sum_{p=\nu+1}^\sdim
	\Wqr_{p\nu}\,\Wgrr_{\kappa\nu p}
	+ \sum_{\sigma = 1}^{\nu-1}(\GSqr^{-1})_{\nu\sigma}
		\WgrW_{\kappa\sigma\nu}
	\r].
	\label{eq:WgraGS}
\end{equation}
Equation~\eqref{WgraGS} is the same as~\eqref{WgrWGS} with~$\mu=\nu$, so that
numerically both~$\WgrW$ and~$\Wgra$ can be obtained in the same loop.

This completes the numerical procedure.  As we mentioned in
\secref{QRbasicmethod}, there is no real additional numerical burden involved
in evaluating~\eqref{WgrWGS} and~\eqref{WgraGS}, as the Lagrangian derivatives
of~$\Wqr$ and~$\nugr_\mu$ are not needed to solve the ODEs.  These derivatives
can be calculated as desired, either at regular intervals or at the end of the
integration.

There are two related but distinct numerical problems when finding the
Lagrangian derivatives.  The first is that the direction of fastest stretching
of the flow dominates and must be isolated from the other directions,
otherwise it quickly becomes impossible to extract subdominant directions
because of lack of numerical precision.  This is the problem we have solved
with our method, by projecting along appropriate characteristic axes.  The
second numerical problem is that the exponentially growing quantities in the
method eventually lead to numerical overflow (or underflow for exponentially
decreasing quantities).  In the \QR\ method for the coefficients of
expansion,~\eqref{nuqrODE} can easily be rewritten as an equation
for~$\log\nuqr_\mu$, replacing the exponential behaviour by linear growth (or
decay) in time.  The same cannot be done in~\eqref{WgrnuODE}--\eqref{WgrrODE}
because the rescaling of the driving term introduces a large damping term that
makes the system extremely stiff (because the rescaling itself is
time-dependent).  But overflowing only becomes a problem if we solve the
system for very long times (on the timescales necessary for the Lyapunov
exponents to converge).

A final note on the stability of the algorithm: in Refs.~\cite{Greene1987}
and~\cite{Christiansen1997} it is shown that the numerical integration of the
orthogonal matrix~$\Qqr$ is unstable (the matrix loses orthogonality), unless
the full spectrum of eigenvalues is calculated, in which case the algorithm is
neutrally stable.  Since we always assume we are computing the full spectrum,
the stability of~$\Qqr$ is not a concern.

\subsection{Numerical Verification of Constraints}
\label{sec:numverconst}

\begin{figure}
\psfrag{t}{$\time$}
\centerline{\includegraphics[width=4.5in]{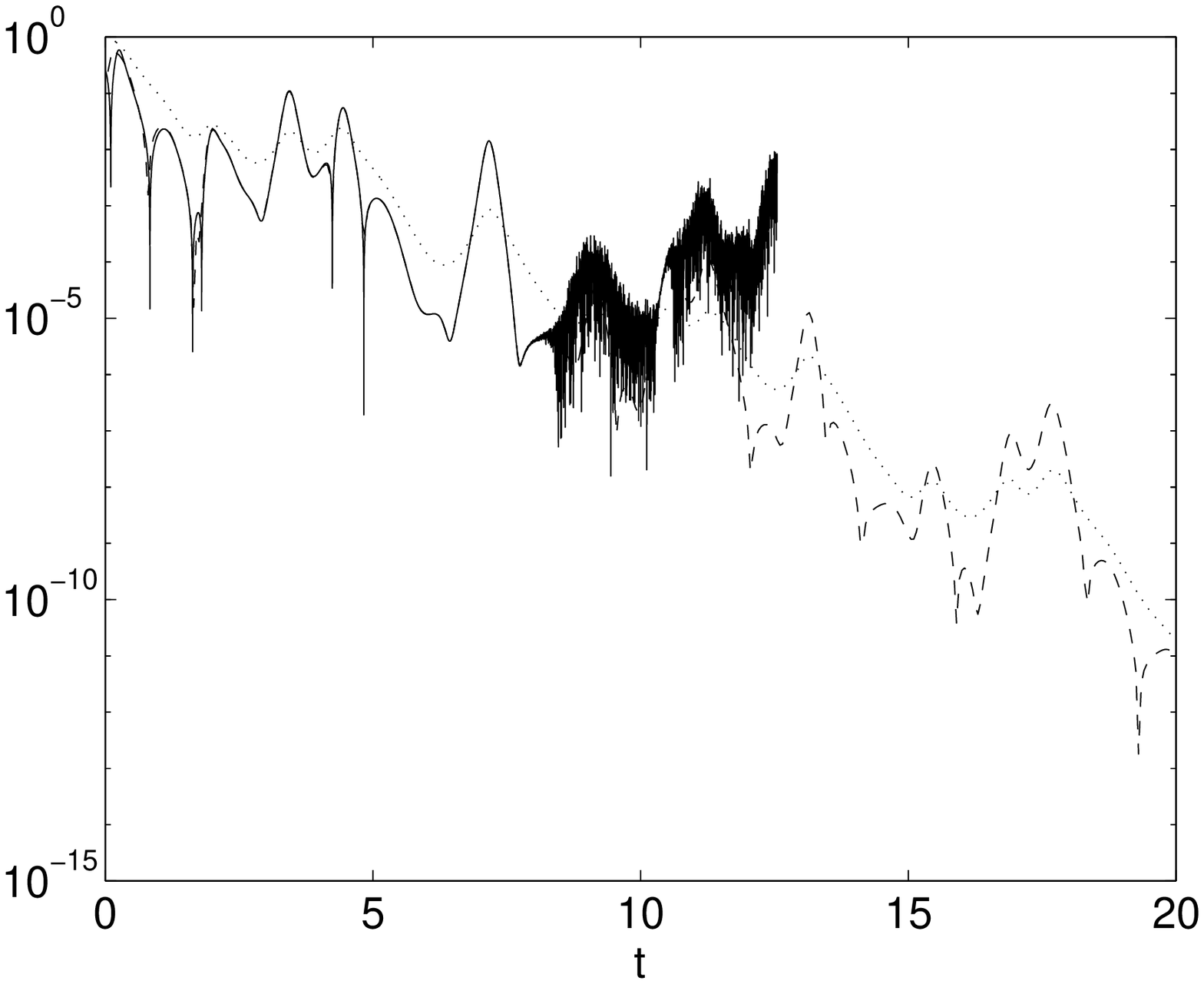}}
\caption{The type I constraint given by~\eqref{divsconstr}, for the \ABC\ flow
with~$A=B=5$, $C=2$, computed with the direct method of \secref{directmethod}
(\full) and with the \QR\ method of \secref{QRmethod} (\broken).  The direct
method becomes unreliable after~$\time\simeq 7$ due to roundoff error.  The
\QR\ method unambiguously exhibits the convergence of the constraint to zero,
which agress well with the predicted convergence rate~$\nugr_3$ (\dotted).}
\label{fig:LogconstrABC522}
\end{figure}

In \figref{LogconstrABC522}, the
type I constraint given by~\eqref{divsconstr} is shown for the \ABC\ flow
with~$A=B=5$, $C=2$~\cite{STF}.  (These parameter values give a large chaotic
region and make the convergence of the constraints faster, but the constraints
are also satisfied for the more usual values~$A=B=C=1$.)  The constraint is
computed with the direct method of \secref{directmethod} and with the \QR\
method of \secref{QRmethod}.  It would be difficult to make a case for the
constraint converging to zero based on the direct method: the noise starting
at~\hbox{$\time\simeq 7$} reflects the effects of limited numerical precision
inherent to the method as the elements of~$\Jacm$ become exponentially large.
The \QR\ method, however, has the constraint reaching~$10^{-12}$ before
precision problems set in (this is not a flaw in the method: the terms
in~\eqref{divsconstr} cannot cancel beyond the number of digits of precision
represented by the machine).  The constraint is predicted
by~\eqref{divsconstr} to converge as~$\nugr_3$, which is also shown in
\figref{LogconstrABC522}.

\begin{figure}
\psfrag{t}{$\time$}
\centerline{\includegraphics[width=4.5in]{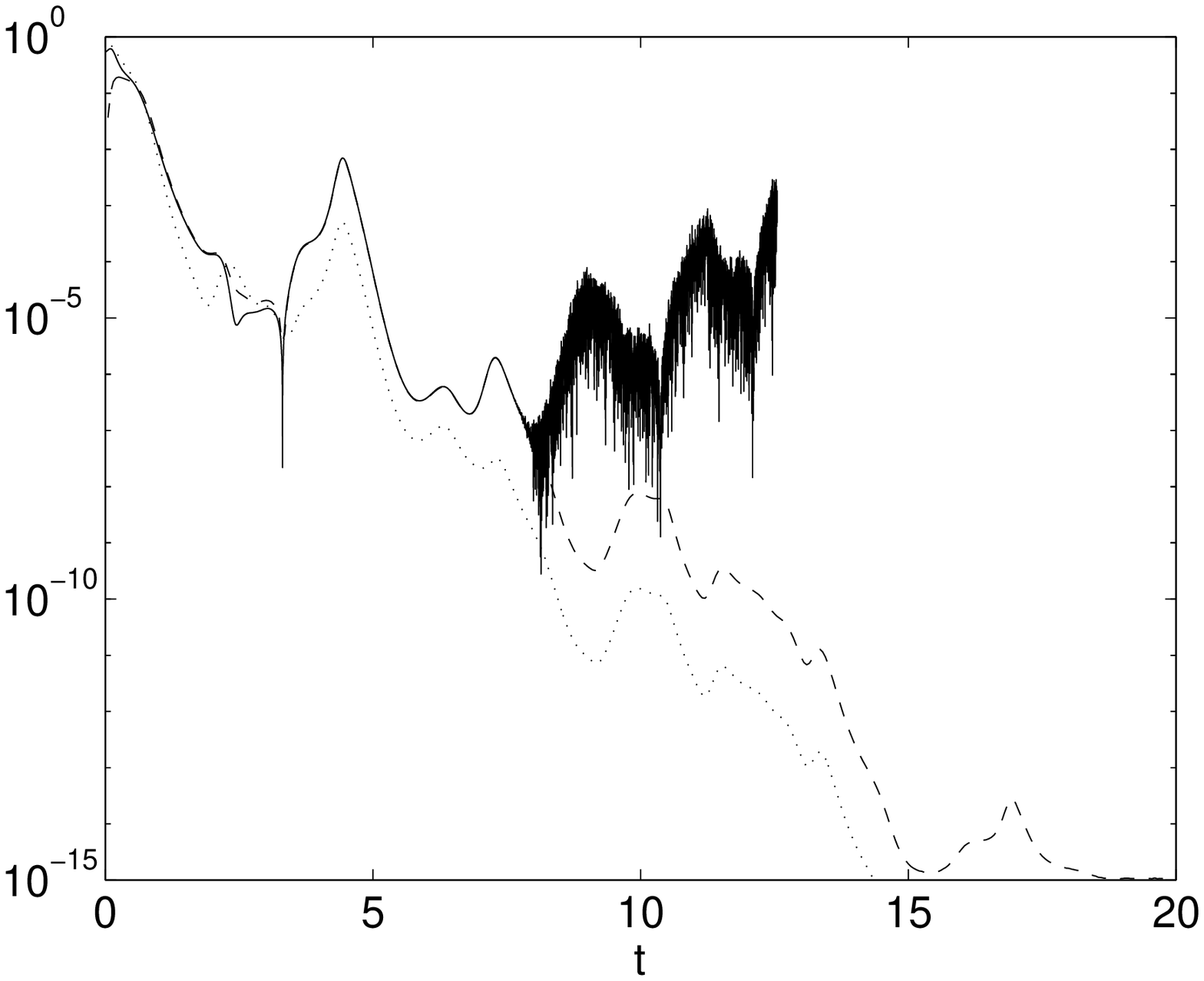}}
\caption{The type II constraint~\hbox{$|\ediruv_1\cdot\curllc\ediruv_1| =
|\VgrV_{231} - \VgrV_{321}|$} [equation~\eqref{constrtype2-3D}] for the \ABC\
flow with~$A=B=5$, $C=2$, computed with the direct method of
\secref{directmethod} (\full) and with the \QR\ method of \secref{QRmethod}
(\dashed).  The direct method becomes unreliable after~$\time\simeq 7$ due to
roundoff error.  The \QR\ method clearly illustrates the convergence
of~$|\ediruv_1\cdot\curllc\ediruv_1|$ to zero, at the predicted rate
of~$(\nugr_2/\nugr_1)^2$ (\dotted).}
\label{fig:LogHeeABC522}
\end{figure}

\Figref{LogHeeABC522} shows a plot of the type II constraint given by
equation~\eqref{constrtype2-3D}, also for the \ABC\ flow.  The constraint
converges as~$(\nugr_2/\nugr_1)^2$, as predicted by~\eqref{constrtype2}.  The
same comments as for \figref{LogconstrABC522} apply regarding numerical
precision.

\section{Discussion}
\label{sec:discussion}

Lagrangian coordinates can greatly simplify the form of partial differential
equations, specifically equations of an advective nature.  We have taken the
viewpoint that to fully characterise quantities expressed in the Lagrangian
frame it is necessary to know how to compute derivatives with respect to these
Lagrangian coordinates.  This amounts to understanding how the Lagrangian
frame itself (as defined by the characteristic eigenvectors and the
coefficients of expansion) vary under small changes of initial conditions.
Obtaining such derivatives with accuracy is difficult in chaotic flows because
the stretching rates of phase-fluid elements vary greatly along different
directions.

The Lagrangian derivatives can be computed by differentiating existing methods
for finding Lyapunov exponents and eigenvectors.  Direct differentiation of
the equations of motion is useful only for short times.  For long times
limited numerical precision becomes problematic and a decomposition method is
needed.  The \SVD\ method proved useful in deriving the asymptotic form of the
Lagrangian derivatives and in deriving differential constraints. The
singularities possessed by the \SVD\ method and its large number of components
make its use in numerical computations difficult.  The \QR\ decomposition
method is more appropriate for numerical implementations, but is less
transparent than the \SVD\ method.  We used the \QR\ method to accurately
verify the differential constraints derived in the paper.

The techniques described here apply only to the first derivatives of the
various quantities, which actually depend on the second derivatives of the
vector field.  First derivatives are sufficient for the study of many systems,
including the advection--diffusion equation and the dynamo problem.  This
seems paradoxical because both the advection--diffusion equation and the
resistive magnetic induction equation involve second derivatives in space, but
when transformed to Lagrangian coordinates the equation only involves first
derivatives of the metric tensor~$\metric$~\cite{Tang1996,Thiffeault2001f}:
the second derivatives are only applied to the initial data.

The results derived in the Lagrangian frame can easily be adapted to the
Eulerian picture with only minor modifications.  The crux of the difference
lies in holding the Eulerian final condition fixed, and integrating the
initial condition backwards in time.  The translation between the Eulerian and
Lagrangian pictures is outlined in \apxref{Eulerian}.

Some of the differential constraints derived in this and earlier papers have
been applied to the study of the advection-diffusion
equation~\cite{Tang1996,Tang1999b,Thiffeault2001f}.  In particular, in
Ref.~\cite{Thiffeault2001f} a type I constraint
(Section~\ref{sec:type1constr}) is used to obtain an effective one-dimensional
diffusion equation for chaotic flows.  In Ref.~\cite{Thiffeault2002b} the
Eulerian form of the type I constraint, equation~\eqref{divuconstr}, is used
to derive a power-law relationship between the curvature of a material line
and the amount of stretching the line has undergone.  This same constraint is
used in~\cite{Giona1998,Adrover1999} to derive an invariant measure of the
spatial length distribution of material lines in two dimensions.  Finally, in
Ref.~\cite{Thiffeault2002c} a type II constraint
(Section~\ref{sec:type2constr}) is used to show that the onset of dissipation
in the kinematic dynamo occurs much later than a straightforward estimate
indicates.  This is because the leading-order behaviour of the power
dissipation (Ohmic heating) in the dynamo is proportional to a type II
constraint, so it does not grow as fast as expected.

The behaviour of second and higher derivatives has not been investigated.
Whilst in principle the method could be extended to cover such cases, the
complexity of the calculation and the smoothness requirements on~$\vel$ are
prohibitive.  A study of the consequences of the degeneracy of Lyapunov
exponents---as occurs for instance in Hamiltonian systems---remains to be
done.

\ack

The author thanks Allen H. Boozer and David Lazanja for helpful discussions.
This work was supported by an NSF/DOE Partnership in Basic Plasma Science
grant, No.~DE-FG02-97ER54441.

\appendix

\section{Asymptotic Behaviour of the Lagrangian Derivatives}
\label{apx:SVDLagrasym}

In this appendix we use the equations of
motion~\eqref{VgrnuODE}--\eqref{VgrVODE} to derive the asymptotic
behaviour~\eqref{VgrUasym}--\eqref{VgrVasym} of the Lagrangian derivatives, as
defined by~\eqref{VgrUnuVdef}.

For~$\mu<\nu$, assuming~$\nugr_\mu\gg\nugr_\nu$, the last term
in~\eqref{VgrUODE} is
\begin{equation}
\fl
	\sum_q\Vsvd_{q\kappa}\,\frac{\pd}{\pd\lagrc^q}
		(\transp{\Usvd}\dot\Usvd)_{\mu\nu}
	= -\sum_q\Vsvd_{q\kappa}\,\frac{\pd\dvdxhat_{\nu\mu}}{\pd\lagrc^q}
	- \sum_q\nudec_{\mu\nu}^2\,
		\Vsvd_{q\kappa}\,\frac{\pd\dvdxhat_{\mu\nu}}{\pd\lagrc^q}
	+ 2\nudec_{\mu\nu}^2\,
		\dvdxsym_{\mu\nu}\,
		(\Vgrnu_{\kappa\mu} - \Vgrnu_{\kappa\nu}).
	\label{eq:lastterm}
\end{equation}
The first term in~\eqref{lastterm} is
\begin{equation}
	\sum_q\Vsvd_{q\kappa}\,\frac{\pd\dvdxhat_{\nu\mu}}{\pd\lagrc^q}
	= \nugr_\kappa\,\Xh_{\nu\kappa\mu}
	+ \sum_\sigma\dvdxhat_{\sigma\mu}\,\VgrU_{\kappa\sigma\nu}
	+ \sum_\sigma\dvdxhat_{\nu\sigma}\,\VgrU_{\kappa\sigma\mu}.
	\label{eq:firsterminlast}
\end{equation}
This shows that the second term in~\eqref{lastterm} can be neglected compared
to the first because it is smaller by a factor~$\nudec_{\mu\nu}^2$.  We also
neglect the third term, which couples~$\VgrU$ and~$\Vgrnu$ (it is
straightforward to go back and check that the neglect is justified).

After these approximations, the evolution equation for~$\VgrU_{\kappa\mu\nu}$
is
\begin{eqnarray}
\fl
\dot\VgrU_{\kappa\mu\nu} =
\sum_\sigma\l[
	(\transp{\Usvd}\dot\Usvd)_{\sigma\mu}\,\VgrU_{\kappa\sigma\nu}
	- (\transp{\Usvd}\dot\Usvd)_{\sigma\nu}\,\VgrU_{\kappa\sigma\mu}
	- \dvdxhat_{\sigma\mu}\,\VgrU_{\kappa\sigma\nu}
	- \dvdxhat_{\nu\sigma}\,\VgrU_{\kappa\sigma\mu}\r]\nonumber\\*
	+ \sum_\sigma
		(\transp{\Vsvd}\dot\Vsvd)_{\sigma\kappa}\,\VgrU_{\sigma\mu\nu}
	- \nugr_\kappa\,\Xh_{\nu\kappa\mu},
	\label{eq:VgrUapprox}
\end{eqnarray}
for~$\mu < \nu$.  This is a linear system in~$\VgrU$, with nonconstant
coefficients and a driving term given by~$-\nugr_\kappa\,\Xh_{\nu\kappa\mu}$.
The only term that couples~$\VgrU_{\kappa\mu\nu}$'s with differing~$\kappa$ is
the~$\transp{\Vsvd}\dot\Vsvd$ one, which is small compared
to~$\transp{\Usvd}\dot\Usvd$.  We neglect this term (again, after the fact it
is easy to check that the neglect is consistent), and
rearrange~\eqref{VgrUapprox} to give
\begin{equation}
\dot\VgrU_{\kappa\mu\nu} =
	(\dvdxhat_{\nu\nu} - \dvdxhat_{\mu\mu})\VgrU_{\kappa\mu\nu}
	+ \sum_{\sigma>\nu}\dvdxsym_{\nu\sigma}\VgrU_{\kappa\mu\sigma}
	- \sum_{\sigma<\mu}
	\dvdxsym_{\mu\sigma}\VgrU_{\kappa\sigma\nu}
	- \nugr_\kappa\,\Xh_{\nu\kappa\mu}.
	\label{eq:VgrUapprox2}
\end{equation}
Let us ignore the driving term for now and consider the homogeneous
solution~$\VgrU^\homo_{\kappa\mu\nu}$.  Through a judicious ordering of
the~$\VgrU_{\kappa\mu\nu}$ that gives the linear part of~\eqref{VgrUapprox2} a
triangular structure, it can be shown that~\hbox{$\VgrU^\homo_{\kappa\mu\nu}
\sim \nudec_{\mu\nu}$}.  If we assume that the motion of the system takes
place in a bounded region of phase space,~$\Xh_{\nu\kappa\mu}$ is also bounded
(we also assume that~$\vel$ is twice differentiable and that its second
derivative is Lipschitz).  Then the inhomogeneous driving
term~$\nugr_\kappa\,\Xh_{\nu\kappa\mu}$ asymptotically goes as~$\nugr_\kappa$.
(A similar argument was used in \secref{SVDbasicmethod} to show convergence
of~$\Vsvd$.)  Asymptotically, then, in~\eqref{VgrUapprox2} either the
exponentially decaying linear part or the driving term dominates, depending on
which has the larger growth rate.  We conclude that
\begin{equation}
	\VgrU_{\kappa\mu\nu} = 
		\max\l(\nugr_\kappa,\nudec_{\mu\nu}\r)
		\VgrUt_{\kappa\mu\nu},
	\qquad \time\gg 1,
	\label{eq:VgrUasym-2}
\end{equation}
where~$\VgrUt_{\kappa\mu\nu}$ is some function that neither grows nor decays
exponentially.

Next, we investigate the asymptotic behaviour of~$\Vgrnu$.  Its time evolution
is given by~\eqref{VgrnuODE}, which after inserting our asymptotic solution
for~$\VgrU$ becomes
\begin{equation*}
\dot\Vgrnu_{\kappa\nu} =
	\sum_\sigma(\transp{\Vsvd}\dot\Vsvd)_{\sigma\kappa}\,\Vgrnu_{\sigma\nu}
	+ 
	\max\l(\nugr_\kappa,\nudec_{\nu\nu}\r)
	\Vgrnudrive_{\kappa\nu},
\end{equation*}
where~$\Vgrnudrive_{\kappa\nu}$ is some non-exponential function.  Notice
that~$\Vgrnu_{\kappa\nu}$'s with differing~$\nu$ are uncoupled.  The
matrix~$\transp{\Vsvd}\dot\Vsvd$ has elements that are decreasing
exponentially, so we can solve the system perturbatively.  The solution, valid
to first-order in~$\transp{\Vsvd}\dot\Vsvd$, is of the form
\begin{equation}
	\Vgrnu_{\kappa\nu} = 
		\max\l(\nugr_\kappa,\nudec_{\kappa\kappa},
		\nudec_{\nu\nu}\r)\Vgrnut_{\kappa\nu}
		+ \Vgrnuinf_{\kappa\nu},
	\qquad \time\gg 1.
	\label{eq:Vgrnuasym-2}
\end{equation}
where the time dependence of~$\Vgrnut_{\kappa\nu}$ is non-exponential.

Finally, having derived an asymptotic form for~$\VgrU$ and~$\Vgrnu$, we can do
the same for the Lagrangian derivatives of~$\Vsvd$, as embodied by~$\VgrV$
[Equation~\eqref{VgrUnuVdef}].  For~$\mu<\nu$,
assuming~$\nugr_\mu\gg\nugr_\nu$, we write~\eqref{VgrVODE} for~$\VgrV$ in the
approximate form
\begin{eqnarray}
\fl
\dot\VgrV_{\kappa\mu\nu} =
\sum_\sigma\l[
	(\transp{\Vsvd}\dot\Vsvd)_{\sigma\kappa}\,\VgrV_{\sigma\mu\nu}
	+ (\transp{\Vsvd}\dot\Vsvd)_{\sigma\mu}\,\VgrV_{\kappa\sigma\nu}
	- (\transp{\Vsvd}\dot\Vsvd)_{\sigma\nu}\,\VgrV_{\kappa\sigma\mu}
\r]\nonumber\\
	+ \nudec_{\mu\nu}\max\l(\nugr_\kappa,\nudec_{\kappa\kappa},
		\nudec_{\mu\mu},\nudec_{\nu\nu}\r)\VgrVdrive_{\kappa\mu\nu},
	\label{eq:VgrVODE2}
\end{eqnarray}
where~$\VgrVdrive_{\kappa\mu\nu}$ is a term with possible time dependence but
without exponential behaviour.  The matrix~$\transp{\Vsvd}\dot\Vsvd$, given
in~\eqref{SVDUVODEapprox}, becomes exponentially small with time.  We can thus
solve~\eqref{VgrVODE2} perturbatively, yielding
\begin{equation}
	\VgrV_{\kappa\mu\nu} =
		\max\l(\nudec_{\mu\nu}\nugr_\kappa,
		\nudec_{\kappa\kappa},
		\nudec_{\mu\mu},
		\nudec_{\nu\nu}\r)
		\VgrVt_{\kappa\mu\nu}
		+ \VgrVinf_{\kappa\mu\nu},
	\qquad \time\gg 1,
	\label{eq:VgrVasym-2}
\end{equation}
where the time dependence of~$\VgrVt_{\kappa\mu\nu}$ is non-exponential.

A comment about the perturbation expansions in small~$\transp{\Vsvd}\dot\Vsvd$
used to obtain~\eqref{Vgrnuasym-2} and~\eqref{VgrVasym-2} is in order.  For a
small parameter~$\varepsilon$, a perturbative expansion solution to an
equation of the form
\begin{equation*}
	\dot y = \varepsilon\, \alpha y + \beta
\end{equation*}
is valid only for~$\varepsilon\,\time \ll 1$.  So for large time, at
fixed~$\varepsilon$, the solution must eventually become invalid.  However, in
our case the parameter~$\varepsilon$, corresponding
to~$\transp{\Vsvd}\dot\Vsvd$, actually decreases exponentially in time (for
nondegenerate eigenvalues~$\nugr_\mu$).  Thus,~$\varepsilon\,\time \ll 1$ even
for large~$\time$, and the expansion remains valid.

\section{The Eigenvalues of~$\metric$ and the \QR\ Method}
\label{apx:QReigensystem}

In this appendix we prove that~\eqref{qreigensystem} gives the asymptotically
correct value of the eigenvectors~$\ediruv_\mu$ and coefficients of
expansion~$\nugr_\mu$ (the square root of the eigenvalues of~$\metric$).  This
result was shown in Ref.~\cite{Goldhirsch1987}.  We present here a different
proof, proceeding by induction and deriving the eigenvectors and eigenvalues
together.

Let~$\GSqr$ be the lower-triangular matrix that performs the Gram--Schmidt
orthonormalisation of~$\rqr$, that is
\begin{equation}
	\transp{\Wqr} = \GSqr\,\rqr,
	\label{eq:GSorthoqrmat}
\end{equation}
where~$\GSqr$ is lower-triangular and~$\Wqr$ is orthogonal (this is
simply~\eqref{GSorthoqr} in matrix form).  The matrix~$\GSqr$ is nonsingular,
so we can invert~\eqref{GSorthoqr} and write~$\rqr = \GSqr^{-1}\transp{\Wqr}$.

Using the definition of the metric,~$\metric = \transp{\Jacm}\Jacm$, and the
\QR\ decomposition~\eqref{QRdecomp}, we have
\begin{equation}
	\metric\,\Wqr = \Wqr\,\DDqr\,,
\qquad
\mathrm{where\ }
\quad
	\DDqr \ldef \transp{(\GSqr^{-1})}\Dqr^2\,(\GSqr^{-1}).
\end{equation}
If~$\DDqr$ were diagonal, then we would be done;  the matrix~$\DDqr$ is not
diagonal, but we show that it can be made so by exponentially small
corrections of the~$\Wqr$.

Writing out~$\DDqr$ explicitly, we find
\begin{equation*}
	\DDqr_{\mu\nu} = \sum_{\sigma=\max(\mu,\nu)}^\sdim
		(\GSqr^{-1})_{\sigma\mu}\nuqr_{\sigma}^2\,
		(\GSqr^{-1})_{\sigma\nu}.
\end{equation*}
The~$\max(\mu,\nu)$ is the lower bound of the sum owing to the
lower-triangular form of~$\GSqr$.  Looking at~\eqref{rqrODE}, it is clear the
the~$\rqr$ converge to constant values, because
for~$\sigma>\mu$,~$\nuqr_\sigma/\nuqr_\mu \rightarrow 0$ exponentially in
time.  Hence, by their definition,~$\Wqr$ and~$\GSqr$ also converge to
constant values in time.  All of the exponential behaviour is thus embodied
in~$\nuqr_{\sigma}$.  Keeping only the dominant term, we have
\begin{equation*}
	\DDqr_{\mu\nu} \simeq
		(\GSqr^{-1})_{\sigma\mu}\nuqr_{\sigma}^2\,
		(\GSqr^{-1})_{\sigma\nu}\Bigr|_{\sigma=\max(\mu,\nu)}.
\end{equation*}
Note that since~$\GSqr$ is triangular we
have~$(\GSqr^{-1})_{\mu\mu}=\GSqr_{\mu\mu}^{-1}$.  For the first column
of~$\Wqr$, it follows that
\begin{equation*}
	\sum_q\metric_{pq}\,\Wqr_{q1} =
	\sum_\sigma\Wqr_{p\sigma}\DDqr_{\sigma1}
	\simeq \frac{\nuqr_{1}^2}{\GSqr_{11}^2}\,\Wqr_{p1}\,,
\end{equation*}
showing that the first column of~$\Wqr$ is an eigenvector of~$\metric$ with
eigenvalue~$(\nuqr_1/\GSqr_{11})^2$.  Let
\begin{equation}
	\Wqr_{p\mu}' = \Wqr_{p\mu} - \sum_{\sigma=1}^{\mu-1}
	\frac{\GSqr_{\sigma\sigma}^2}{\nuqr_\sigma^2}\,\DDqr_{\mu\sigma}\,
	\Wqr_{p\sigma},
	\label{eq:Wcorr}
\end{equation}
which represents an exponentially small correction to the matrix
element~$\Wqr_{p\mu}$,
because~\hbox{$\DDqr_{\mu\sigma}\sim\nuqr_\mu^2\ll\nuqr_\sigma^2$}
for~\hbox{$\sigma<\mu$}.  Assume that the columns~$\Wqr_{q\nu}'$, $\nu<\mu$,
are eigenvectors of~$\metric$ with eigenvalue~$(\nuqr_\nu/\GSqr_{\nu\nu})^2$.
We show by induction that with the (small) correction~\eqref{Wcorr} the
column~$\Wqr_{q\mu}'$ is an eigenvector of~$\metric$ with
eigenvalue~$(\nuqr_\mu/\GSqr_{\mu\mu})^2$.

Using~\hbox{$\DDqr = \transp{\Wqr}\metric\,\Wqr$} and~\eqref{Wcorr}, the
corrected matrix element~$\DDqr_{\mu\nu}'$, with~$\nu<\mu$, is
\begin{equation*}
\DDqr_{\mu\nu}' = \Wqr_{p\mu}'\,\metric_{pq}\,\Wqr_{q\nu}
	= \DDqr_{\mu\nu} - \sum_{\sigma=1}^{\mu-1}
	\frac{\GSqr_{\sigma\sigma}^2}{\nuqr_\sigma^2}\,
	\DDqr_{\mu\sigma}\,\DDqr_{\sigma\nu}, \qquad \nu<\mu.
\end{equation*}
We use the induction hypothesis that~$\DDqr_{\sigma\nu} =
(\nuqr_\nu/\GSqr_{\nu\nu})^2\,\delta_{\sigma\nu}$ when both~$\nu$ and~$\sigma$
are less than~$\mu$, and find
\begin{equation*}
	\DDqr_{\mu\nu}' = \DDqr_{\mu\nu} - \sum_{\sigma=1}^{\mu-1}
	\frac{\GSqr_{\sigma\sigma}^2}{\nuqr_\sigma^2}\,
	\DDqr_{\mu\sigma}\,\frac{\nuqr_\nu^2}{\GSqr_{\nu\nu}^2}
	\,\delta_{\sigma\nu}
	= \DDqr_{\mu\nu} - \DDqr_{\mu\nu} = 0,
\end{equation*}
to exponential accuracy.  The corrected diagonal element~$\DDqr_{\mu\mu}'$ is
\begin{eqnarray}
	\DDqr_{\mu\mu}' &= \DDqr_{\mu\mu} - 2\sum_{\sigma=1}^{\mu-1}
	\frac{\GSqr_{\sigma\sigma}^2}{\nuqr_\sigma^2}\,
	\DDqr_{\mu\sigma}\,\DDqr_{\sigma\mu}
	+ \sum_{\sigma=1}^{\mu-1}\sum_{\tau=1}^{\mu-1}
	\frac{\GSqr_{\sigma\sigma}^2}{\nuqr_\sigma^2}\,
	\frac{\GSqr_{\tau\tau}^2}{\nuqr_\tau^2}\,
	\DDqr_{\mu\sigma}\,\DDqr_{\mu\tau}\,\DDqr_{\sigma\tau}\nonumber\\
&= \DDqr_{\mu\mu} - \sum_{\sigma=1}^{\mu-1}
	\frac{\GSqr_{\sigma\sigma}^2}{\nuqr_\sigma^2}\,
	\DDqr_{\mu\sigma}^2
\simeq \DDqr_{\mu\mu}
\label{eq:DDcorr}
\end{eqnarray}
to leading order.  Thus, to exponential accuracy~$\Wqr_{q\mu}'$ is indeed an
eigenvector of~$\metric$ with eigenvalue~$\DDqr_{\mu\mu} =
(\nuqr_\mu/\GSqr_{\mu\mu})^2$.

To complete the proof, we need to show that the columns~$\Wqr_{q\nu}$,
with~$\nu>\mu$, are not modified by the correction~\eqref{Wcorr}.  We have
\begin{equation*}
	\DDqr_{\mu\nu}' = \DDqr_{\mu\nu} - \sum_{\sigma=1}^{\mu-1}
	\frac{\GSqr_{\sigma\sigma}^2}{\nuqr_\sigma^2}\,
	\DDqr_{\mu\sigma}\,\DDqr_{\sigma\nu}, \qquad \nu>\mu,
\end{equation*}
which to leading order is
\begin{equation*}
	\DDqr_{\mu\nu}' =
	(\GSqr^{-1})_{\nu\mu}\,(\GSqr^{-1})_{\nu\nu}\,\nuqr_{\nu}^2 +
	\Order{\frac{\nuqr_\mu^2}{\nuqr_{\mu-1}^2}\,\nuqr_\nu^2}
	\simeq \DDqr_{\mu\nu},
\end{equation*}
showing that the correction can be neglected.

We have thus proved by induction that the correction~\eqref{Wcorr} to~$\Wqr$
makes~$\DDqr$ diagonal to leading order, leaving its diagonal elements
unaffected.  However, the correction~\eqref{Wcorr} is of
order~$(\nuqr_\mu/\nuqr_{\mu-1})^2$, which is exponentially small with time.
We conclude that the eigenvectors of~$\metric$ and corresponding coefficients
of expansion are
\begin{equation*}
	(\ediruv_{\mu})_q = \Wqr_{q\mu}, \qquad
	\nugr_\mu = \frac{\nuqr_\mu}{\GSqr_{\mu\mu}}\,,
\end{equation*}
to exponential accuracy with time.  The relative error on~$\ediruv_\mu$
and~$\nugr_\mu^2$ is of order~$(\nuqr_\mu/\nuqr_{\mu-1})^2$, as can be seen
from the leading-order correction in~\eqref{Wcorr} and~\eqref{DDcorr}.

\section{The Eulerian Perspective}
\label{apx:Eulerian}

The results derived in the paper regarding Lagrangian derivatives can readily
be adapted to an Eulerian framework.  The Lagrangian derivatives can be
regarded as measuring the effect of an infinitesimal change in the initial
condition of a trajectory.  Conversely, one can regard Eulerian derivatives as
the effect of an infinitesimal change in the \emph{final} condition of a
trajectory, with the integration being performed backwards in time.  This is
the viewpoint taken in studies of the alignment of material lines with the
Eulerian unstable manifold of a system, a phenomenon referred to as
\emph{asymptotic
directionality}~\cite{Cerbelli2000,Giona1998,Adrover1999,Giona1999,Giona2000}.

In our framework, the Eulerian characteristic directions are computed from the
metric
\begin{equation}
	\metricx^{ij}(\time,\time_0,\xv) \ldef
		\sum_{p=1}^\sdim{\Jacm^i}_p\,{\Jacm^j}_p,
	\label{eq:metricxdef}
\end{equation}
or~$\metricx\ldef\Jacm\transp{\Jacm} = \transp{\metric}$.  We have explicitly
written the dependence on initial time because we are now interested in
evolving $\time_0$ towards~$-\infty$, whilst holding the Eulerian
coordinates~$\time$ and~$\xv$ fixed.  The dynamical system~\eqref{dynsys} and
the \SVD\ equations~\eqref{SVDnugrODE}--\eqref{SVDVODE} are also evolved
backwards in time: the method yields~$\Jacm^{-1}$, so that we must take the
inverse of the resulting~$\nugr_\sigma$ to obtain the ``forward-time''
coefficients of expansion.  We assume the forward-time coefficients have then
been reordered in the usual decreasing manner, so that~$\nugr_1$ is still the
fastest-growing coefficient (the columns of~$\Usvd$ and~$\Vsvd$ are also
reordered).  The asymptotic behaviour of the Eulerian derivatives will thus be
the inverse of their Lagrangian counterpart.  The columns of~$\Vsvd$ now
contain vectors associated with the Eulerian frame.
%
%
The relevant Eulerian definitions corresponding to~\eqref{VgrUnuVdef} simply
involve replacing~$\pd/\pd\lagrcv$ by~$\pd/\pd\xv$ to reflect the fact the the
derivatives are now taken with respect to the Eulerian coordinates.  Their
asymptotic behaviour is
\begin{eqnarray}
\UgrV_{\kappa\mu\nu} =
		\max\l(\nugr_\kappa^{-1},\nudec_{\mu\nu}\r)
		\UgrVt_{\kappa\mu\nu},
		\label{eq:UgrVasym}\\
\Ugrnu_{\kappa\nu} =
		\max\l(\nugr_\kappa^{-1},\nudec_{\kappa\kappa},
		\nudec_{\nu\nu}\r)\Ugrnut_{\kappa\nu}
		+ \Ugrnuinf,
		\label{eq:Ugrnuasym}\\
\UgrU_{\kappa\mu\nu} =
		\max\l(\nudec_{\mu\nu}\nugr_\kappa^{-1},
		\nudec_{\kappa\kappa},
		\nudec_{\mu\mu},
		\nudec_{\nu\nu}\r)
		\UgrUt_{\kappa\mu\nu}
		+ \UgrUinf,
	\label{eq:UgrUasym}
\end{eqnarray}
where the~${}^\Eul$ superscript reminds us that these are Eulerian quantities.
The discussion of the Hessian and constraints in \secref{symhesconstraints} is
the same for the Eulerian derivatives, except that the coefficients of
expansion corresponding to contracting directions are replaced by the inverse
of those of the expanding directions.  For instance, the type I constraint
given by~\eqref{divsconstr} becomes
\begin{equation}
\fl
	\sum_{i}\detmetricx^{1/2}\,
	\frac{\pd}{\pd\x^i}\l(\detmetricx^{-1/2}\,(\ediruv^\Eul_{1})_i\r)
	+ \sum_i(\ediruv^\Eul_{1})_i\frac{\pd}{\pd\x^i}\log\nugr_1
	\sim \max\l(\nugr_1^{-1}\,,\,\nudec_{11}^2\r)
	\rightarrow 0.
	\label{eq:divuconstr}
\end{equation}
This is a more general form of the relation derived for the 2D incompressible
case in Refs.~\cite{Giona1998,Adrover1999}, used to derive an invariant
measure of the spatial length distribution of material lines.

\section*{References}


\end{document}